\documentclass[usenatbib,onecolumn]{mn2e}
\usepackage{newtxtext,newtxmath}
\usepackage{lipsum}
\usepackage[T1]{fontenc}
\usepackage{ae,aecompl}
\usepackage{hyperref}
\usepackage{graphicx}	% Including figure files
\usepackage{amsmath}	% Advanced maths commands
\usepackage{amssymb}	% Extra maths symbols
\usepackage{bm}	% Extra maths symbols
\usepackage{caption}
\usepackage{multirow}
\usepackage{multicol}
\usepackage{cuted}
\usepackage{mathtools}
\usepackage{longtable}

\title[Nearest Jupiter analog]{Detection of the nearest Jupiter analog in radial velocity and astrometry data}
\author[F. Feng et al.]
{Fabo Feng$^{1}$\thanks{E-mail: fengfabo@gmail.com},
  Guillem Anglada-Escud\'e$^{2}$, Mikko Tuomi$^{3}$, Hugh R. A.  Jones$^{3}$,
  \newauthor
Julio Chanam\'e$^{4,5}$, Paul R. Butler$^{1}$, Markus Janson$^{6}$ \\
$^{1}$Department of Terrestrial Magnetism, Carnegie Institution of
Washington, 5241 Broad Branch Road, NW, Washington, DC 20015, USA\\
$^{2}$School of Physics and Astronomy, Queen Mary University of London, 327 Mile End Road, London, United Kingdom\\
$^{3}$Centre for Astrophysics Research, University of Hertfordshire, College Lane, AL10 9AB, Hatfield, UK \\
$^{4}$Instituto de Astrof\'isica, Pontificia Universidad Cat\'olica 
de Chile, Av.~Vicu\~na Mackenna 4860, 782-0436 Macul, Santiago, Chile\\
$^{5}$Millennium Institute of Astrophysics, Santiago, Chile\\
$^{6}$Department of Astronomy,Stockholm University, AlbaNova
University Center, 10691 Stockholm, Sweden\\
}
\date{\today}

\begin{document}
\maketitle
\begin{abstract}
  The presence of Jupiter is crucial to the architecture of the Solar
  System and models underline this to be a generic feature of
  planetary systems. We find the detection of the difference between
  the position and motion recorded by the contemporary astrometric
  satellite Gaia and its precursor Hipparcos can be used to discover
  Jupiter-like planets. We illustrate how observations of the nearby
  star $\epsilon$ Indi A giving astrometric and radial velocity data can be used to independently find the orbit of its suspected companion. The radial velocity and astrometric data provide complementary detections which allow for a much stronger solution than either technique would provide individually. We quantify $\epsilon$
  Indi A b as the closest Jupiter-like exoplanet with a mass of 3
  $M_{\rm Jup}$ on a slightly eccentric orbit with an orbital period of 45
  yr. While other long-period exoplanets have been discovered, $\epsilon$ Indi
  A b provides a well constrained mass and along with the well-studied
  brown dwarf binary in orbit around $\epsilon$ Indi A means that the system
  provides a benchmark case for our understanding of the formation of
  gas giant planets and brown dwarfs.
\end{abstract}
%%%%%%%%%%%%%%%%%%%%%%%%%%%%%%%%%%%%%%%%%%%%%%%%%%%%%%%%%%%%%%%%%%%%%%%%%%%%
\section{Introduction}     \label{sec:introduction}
One of the main goals of exoplanet searches is to understand our Solar
System, with inner and outer planets including the benchmark planets:
Earth and Jupiter. Although hot Jupiters around Sun-like stars and
Earth-mass planets around M-dwarfs have been detected during the last
decades \citep{gillon16,anglada-escude16} few Jupiter-like planets have been detected around
Sun-like stars and such detections when they are made have only been
possible with a single technique (e.g., \citealt{kuzuhara13,wittenmyer16}).

$\epsilon$ Indi A is a good candidate to search for Solar System
analogs. $\epsilon$ Indi A (HIP 108870, HR 8387, HD 209100, GJ 845) is
a nearby K2V star (3.62\,pc according to \citealt{leeuwen07}) with a
mass of 0.762$\pm$0.038\,$M_\odot$ \citep{demory09}, and a luminosity
of 0.22\,$L_\odot$. This star is also accompanied by a binary brown
dwarf with a separation of about 1459\,au \citep{scholz03}. A clear long-term signal has been established in the radial velocity data by \cite{endl02} and \cite{zechmeister13}. This signal is much larger than from the relatively distant brown dwarfs and supports the existence of another companion with a period longer than 30 years. The non-detection of this signal in direct imaging by \cite{janson09} suggests that the companion inducing the radial velocity signal has a relatively low temperature and a particularly interesting object for follow-up by multiple techniques. 

To find small Keplerian signals and to investigate the previously proposed long period companion around $\epsilon$ Indi A, we analyze the data from \cite{zechmeister13} and the recent HARPS data in the ESO archive in the Bayesian framework. With new data and noise modelling techniques there is the potential for detection of weak signals (e.g. \citealt{feng16, feng17b}). Using the {\small Agatha} software \citep{feng17a}, we compare noise models to find the so-called Goldilocks model \citep{feng16}. We also calculate the Bayes factor periodograms (BFPs) to test the sensitivity of signals to the choice of noise models. These signals are further diagnosed by calculating the BFPs for various noise proxies and their residuals.

The combination of RV and astrometry has been used to detect and
characterize a few short period planets, though no cold Jupiters have
been detected in a similar way. Thanks to the two decades of baseline
provided by the Hipparcos and Gaia survey
\citep{leeuwen07,brown18,lindegren18}, we now have at least two epochs
with well determined positions and proper motions, which are quite sensitive to the stellar reflex motion
caused by Jupiter-like planets. Since the barycentric motion of the
satellite is modeled in combination with proper motions in the
analyses of the raw data \citep{lindegren18}, the astrometric survey data provides the
position and velocity of a star at a given epoch viewed from the
barycenter of the Solar System. $\epsilon$ Indi A provides a test case for a nearby object with a significant change in RV and change in position and proper motion between Hipparcos and Gaia over a long time baseline of observations.

This paper is structured as follows. First, we introduce the data in
section \ref{sec:data}. Then we describe the statistical and numerical
methods for the analysis of RV data and constrain the long period
signal present in the data. In section \ref{sec:signal}, we
concentrate on the HARPS data and constrain the activity signals using
posterior samplings. In section \ref{sec:combined}, we analyze the RV
and astrometry data to constrain the mass and orbital parameters of
$\epsilon$ Indi A b. Finally, we discuss and conclude in section \ref{sec:conclusion}. 

\section{RV data and model}\label{sec:data}
We obtain the HARPS data by processing the spectra in the ESO archive
(Programmes 60.A-9036, 072.C-0488, 072.C-0513, 073.C-0784, 074.C-0012,
077.C-0530, 078.C-0833, 079.C-0681, 081.D-0870, 183.C-0972,
087.D-0511, 091.C-0853, 091.C-0844, 094.C-0894, 192.C-0852,
196.C-1006, 098.C-0446) using the TERRA algorithm
\citep{anglada12}. The HARPS data also include various noise proxies
courtesy of the HARPS pipeline, including Bisector span (BIS) and full
width half maximum of spectra lines (FWHM). These are supplemented by
TERRA-generated indices including R$'_{\rm HK}$ (or CaHK index),
intensity of the H-alpha line (H-alpha), indices from sodium lines
(NaD1 and NaD2), and the differential RV sets which record
  wavelength dependent noise \citep{feng17b}. We only use the 3AP3-2
  differential set which is found to be correlated with RVs.

\begin{figure*}
  \begin{center}
\includegraphics[scale=0.55]{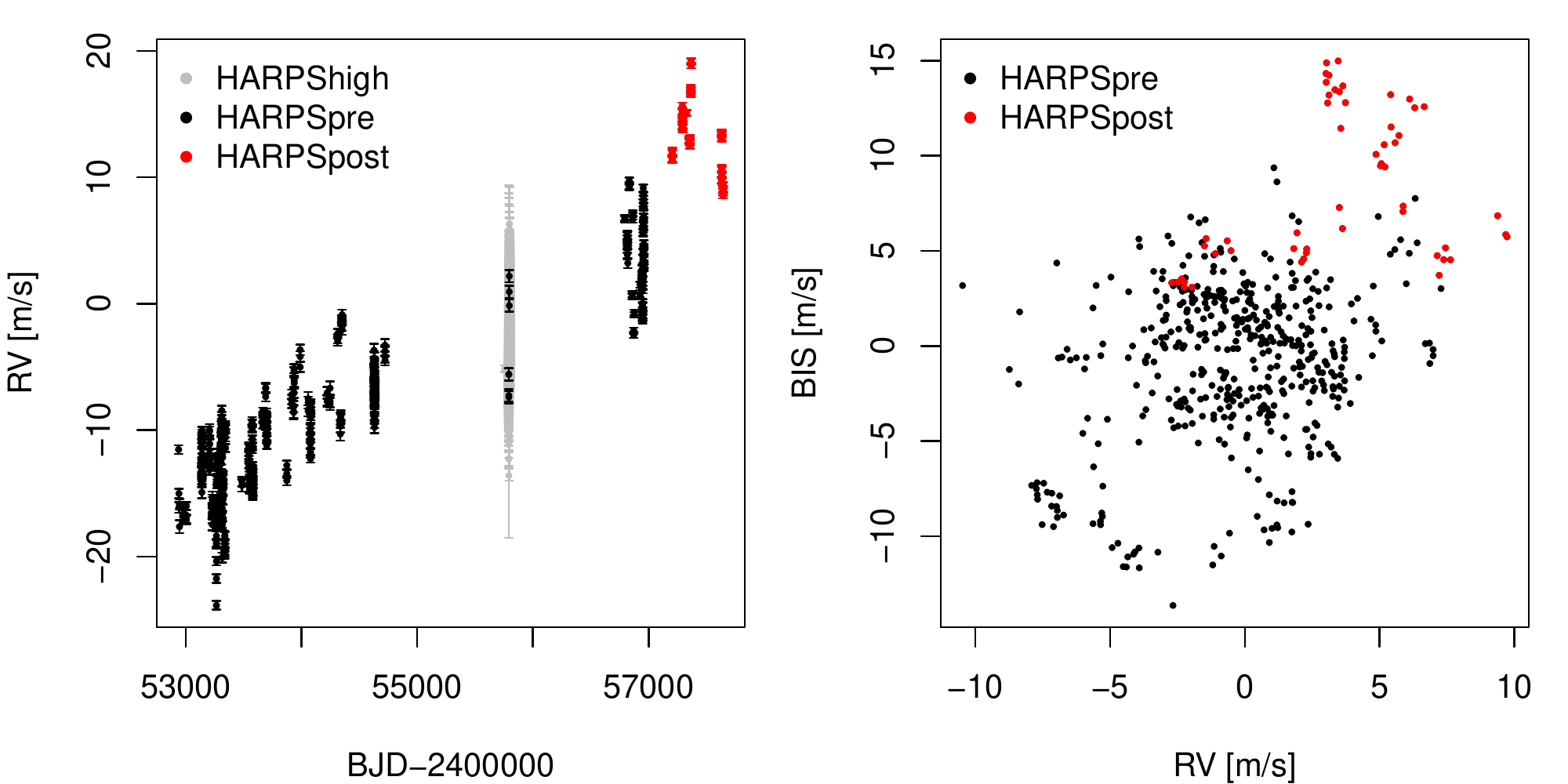}
\caption{HARPS data of $\epsilon$ Indi A. Left panel: the HARPS data set consists of the high cadence RVs (HARPShigh) with signal to noise ratio less than 110 (grey) and the low cadence data (HARPSlow). The HARPSlow set includes the pre-2015 (HARPSpre; black) and post-2015 (HARPSpost; red) RVs. Right panel: correlation between BISs and RVs of the HARPSlow set after subtraction by corresponding best-fitting parabola.}
\label{fig:data}
\end{center}
\end{figure*}

In the left-hand panel of Fig. \ref{fig:data}, we show all the HARPS RVs including high cadence epochs (called ``HARPS''; 4198 points). We note that the RVs from JD2455790 to JD2455805 are measured with high cadence, leading to 3636 RVs spread over two weeks. Such high-cadence data were obtained to study high frequency stellar oscillations and most of them are measured with a considerably lower signal to noise ratio than the rest of the data. To remove the stellar and instrumental systematics, we define the HARPSlow data set by excluding high cadence RVs with signal to noise ratio less than 110 (HARPShigh; shown by the grey points in Fig. \ref{fig:data}). This dataset consists of 518 points. We are aware of the ill-determined offset for post-2015 data \citep{curto15} due to fibre change, which also give relatively high BIS (see the right panel of Fig. \ref{fig:data}). We name the pre-2015 data set as ``HARPSpre'', and the post-2015 data as ``HARPSpost''. Thus the HARPS set is a combination of HARPSlow and HARPShigh while the HARPSlow is a combination of HARPSpre and HARPSpost. We will use these sub-sets to test the sensitivity of signals to the choice of datasets\footnote{All of these data sets are available at \url{http://star-www.herts.ac.uk/~ffeng/research\_note/eps_indi/}.} in section \ref{sec:signal}. 

An offset parameter is needed to combine HARPSpre and HARPSpost. To
avoid potential degeneracy between this offset and Keplerian signals
especially with period comparable with the HARPSpost time span, we use
HARPSpre to investigate short-period signals. To constrain the
  long period signal, we use HARPSpre, HARPSpost as well as the CES
  long camera (LC) and very long camera (VLC) data sets from
  \cite{zechmeister13}. We also use the UVES/VLT RVs which are
  measured at epochs of 2453272, 2457905, and 2457993 to constrain the
  trend. The UVES data is shown in Tables \ref{tab:data}, \ref{tab:data2},
  and \ref{tab:data3}. The other data sets are available at
    \url{http://star.herts.ac.uk/~hraj/HD209100/}.

To model the RV variation caused by stellar activity and
  planetary perturbations, we use a combined moving average (MA) and Keplerian model:  
\begin{equation}
  \hat{v}_r(t_i)=v_r^p(t_i)+\sum_{l=1}^q w_l {\rm
    exp}\left(-\frac{|t_i-t_{i-k}|}{\tau}\right)~,
  \label{eqn:rv1}
\end{equation}
where $\{w_l\}$ and $\tau$ are respectively the amplitude and
timescale of the $q^{\rm th}$ order MA model (or MA(q)). The
planet-related RV variation is
\begin{equation}
  v_r^p(t_i)=\sum_{k=1}^{N_p}v_k^p(t_i)+b~,
    \label{eqn:rv2}
\end{equation}
where $b$ is the RV offset and independent offsets are adopted for
different instruments. The RV variation caused by planet $k$ is
\begin{equation}
  v_k^p(t_i)=\sqrt{\frac{Gm_{pk}^2}{(m_{pk}+m_s)a_k(1-e_k^2)}}\sin{I_k}\left[\cos{\omega_k+\nu_k(t_i)}+e_k\cos(\omega_k)\right]~,
      \label{eqn:rv3}
\end{equation}
where $m_{pk}$ is the mass, $a_k$, $e_k$, $I_k$, $\omega_k$  and
$\nu_k$ are the orbital elements of the $k^{\rm th}$ planet. The true anomaly
and eccentricity anomaly are derived from the mean anomaly at the
reference epoch for a given time. Independent sets of parameters of
the MA model are applied to different data sets. We use a superscript to
denote the parameter for a given set. For example, $\tau^{\rm HARPSpre}$
is the correlation time scale for the HARPSpre data set.

\section{Confirmation of a cold Jupiter}\label{sec:detection}
A significant RV trend has been found in the CES and HARPS data sets
\citep{endl02,zechmeister13}, suggesting a long-period planetary
companion to $\epsilon$ Indi A. To constrain this RV trend better by
including newer HARPS and UVES data, we analyze the CES LC and VLC
data sets in combination with the HARPSpre,  HARPSpost, and UVES
sets. The LC and VLC data sets are corrected by accounting for the
1.8\,m\,s$^{-1}$/\,yr acceleration caused by the proper motion (so-called ``perspective acceleration''). Based on Bayesian model comparison \citep{feng17a}, we model the trend in the combined data using one Keplerian function, and model the noise in LC, VLC, and UVES using white noise. We use the first and second order moving average models (i.e. MA(1) and MA(2)) respectively to model the noise in HARPSpre and HARPSpost, in the same manner as e.g., \cite{tuomi12,feng16}.

We also vary the offsets between data sets and adopt the prior
distributions for all parameters from \cite{feng16}. Specifically, we
use a one-sided Gaussian prior distribution $\mathcal{N}(0,0.2)$ for
eccentricity to account for the eccentricity distributions found
  in radial velocity planets \citep{kipping13} and in transit systems
  \citep{kane12,VanEylen18}. Since there is no universal eccentricity
  distribution for all types of planets and stars, our use of a semi-Gaussian
  only captures the broad feature of the real eccentricity
  distributions. Nevertheless, we will test the sensitivity of our
  results to eccentricity priors later. We use a log uniform distribution for time scale parameters, and a uniform distribution for other parameters. Based on adapted MCMC samplings \citep{haario06}, we find a significant signal at a period of $17155_{-4748}^{+6940}$\,d or $47_{-13}^{+19}$\,yr, which is well constrained by MCMC samplings. The one-planet model improves the maximum likelihood by about $10^{144}$ (or BF=$10^{136}$) compared with the baseline model. It increases the maximum likelihood by about $10^{37}$ (or BF=$10^{31}$) compared with the baseline model combined with a linear trend, and increase the maximum likelihood by $10^{22}$ (or BF=$10^{18}$) compared with the baseline combined with a parabola. Hence a linear trend or parabola or a longer period signal is not favored by the data.

\begin{figure}
 \begin{center}
\includegraphics[scale=0.6]{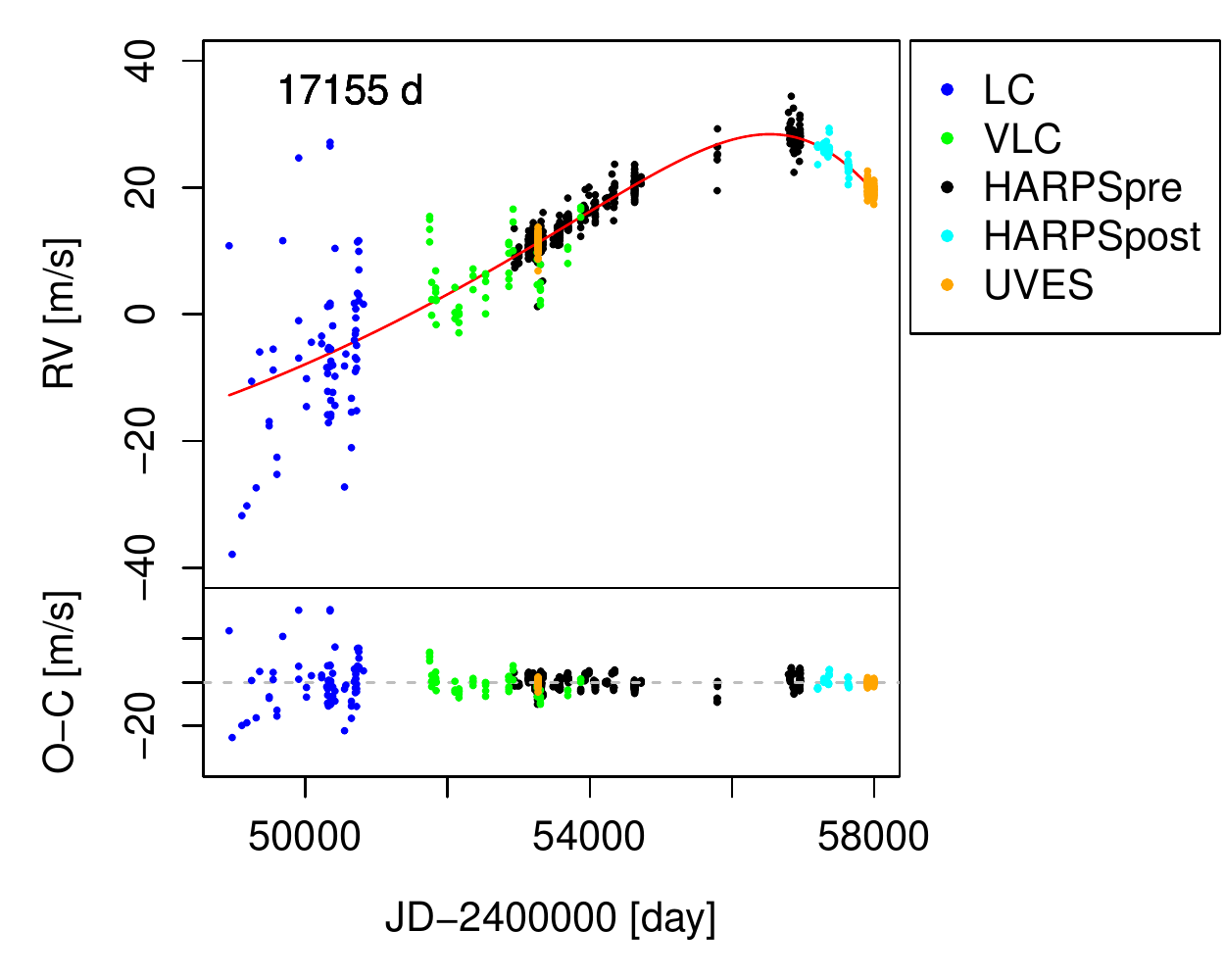}
 \caption{Best-fit for the data sets of CES-LC, CES-VLC, HARPSpre,
   HARPSpost, and UVES. The offset and correlated noise are subtracted
   from the data. The red line denotes the Keplerian signal with a
   period of 47\,yr based on the analysis of RV data. }
\label{fig:fit}
\end{center}
\end{figure}

Since this trend is not found in noise proxies, it is likely to be
caused by a wide-orbit planet with a minimum mass $m\sin{I}=2.6_{-0.72}^{+2.0}~M_{\rm Jup}$, semi-major axis
  $a=12_{-2.0}^{+2.1}$\,au, semi-amplitude
  $K=25_{-5.5}^{+16}$\,m/s, eccentricity of
  $e=0.26_{-0.071}^{+0.078}$, argument of periapsis
  $\omega=60_{-16}^{+35}$\,deg, and mean anomaly of
  $M_0=170_{-82}^{+50}$\,deg at the first epoch (or
    BJD2448929.56) of the combined RV data. The optimal value is determined at the {\it maximum a posteriori} and the uncertainties correspond to the 10 and 90\% quantiles of the posterior distributions. This solution
for $\epsilon$ Indi Ab is shown in Fig. \ref{fig:fit}. The offset
between VLC and LC sets is 4.6$\pm$2.8\,m/s, which is within the
uncertainty of 8\,m/s determined from a sample of VLC and LC sets
\citep{zechmeister13}. The offset between the HARPSpre and HARPSpost
set is 12$\pm$1.3\,m/s, consistent with the offset between
pre-and-post fibre exchange for a K2 star given by \cite{curto15}. The
mass determined in this work is consistent with the range given in
\cite{janson09}, who adopt a 2.6\,m\,s$^{-1}$/\,yr planet-induced
acceleration of $\epsilon$ Indi A relative to its barycenter (or
stellar reflex motion) determined from epochs earlier than JD2455000
(or June, 2009), leading to an estimation of mass limit of 5-20$M_{\rm Jup}$. On the other hand, \cite{zechmeister13} adopt a 2.4\,m\,s$^{-1}$/\,yr slope and find a broad limit of $M\sin{i}>0.97$\,$M_{\rm Jup}$ and
  $P>30$\,yr. With the benefit of the number of years of precise
HARPS epochs we are in a position to determine a more modest
slope value of 1.8\,m\,s$^{-1}{\rm yr}^{-1}$ for HARPSpre
epochs and a negative slope value of -2.9\,m\,s$^{-1}{\rm yr}^{-1}$ for the HARPSpost epochs. Moreover, the combination of UVES and HARPS sets also strongly favor a curvature in the RV variation. The early epochs of UVES overlap with HARPS epochs and thus enable a good offset calibration. The recent epochs of UVES suggest a significant decrease in RV (see Fig. \ref{fig:fit}) as already seen in the HARPSpost set. Thus a Keplerian function is more appropriate than the previously used linear trend to model the RV variation and constrain the signal. 

Since $\epsilon$ Indi Ab is on such a very wide orbit and so
  within the dataset of known planets can not be expected to follow
  the eccentricity distribution found in relatively short period
  planets \citep{kipping13,juric08}. Hence we test whether the values
  of orbital parameters are sensitive to the choice of eccentricity
  prior. We change the standard deviation of the one-sided Gaussian
  prior from 0.2 to 0.4, and also adopt a uniform eccentricity
  prior. The orbital parameters for different parameters and the log
  BF are shown in Table \ref{tab:test}. It is evident that the orbital
  parameters from different solutions and the offsets as well as
    jitters are consistent with each other and are thus not sensitive to the choice of eccentricity prior. 
  \begin{table}
    \centering
    \caption{Sensitivity of orbital parameters to the choice of
      eccentricity priors. The optimal value is calculated at the {\it
        maximum a posteriori} (MAP) and the uncertainties correspond
      to the 10\% and 90\% qantiles. The ln(BF) of one-planet model
      with respect to the baseline model is estimated by the Bayesian
      information criterion (BIC; \citealt{schwarz78}), and ln(BF)>5
      is the signal detection threshold \citep{feng17a}. }
    \renewcommand{\arraystretch}{1.25}
    \begin{tabular}{l*3{c}}
      \hline\hline
      & $P(e)=2\mathcal{N}(0,\,0.2)$ &$P(e)=2\mathcal{N}(0,\,0.4)$&
                                                           $P(e)=1$\\
      &$0<e<1$&$0<e<1$&$0<e<1$\\\hline
      $P$ [yr]&$47_{-13}^{+19}$&$44_{-9.3}^{+23}$&$49_{-14}^{+19}$\\
      $K$ [m/s]&$25_{-5.5}^{+16}$&$29_{-9.1}^{+13}$&$28_{-7.9}^{+1.7}$\\
      $e$&$0.26_{-0.071}^{+0.078}$&$0.26_{-0.049}^{+0.11}$&$0.26_{-0.054}^{+0.11}$\\
      $\omega$ [deg]&$60_{-16}^{+35}$&$80_{-36}^{+17}$ &$64_{-18}^{+32}$\\
      $M_0$ [deg]&$170_{-82}^{+50}$ &$140_{-54}^{+80}$&$167_{-80}^{+51}$\\
      $b^{\rm LC}$&$-17_{-15}^{+7.7}$&$-17_{-16}^{+7.7}$&$-19_{-15}^{+9.7}$\\
      $\sigma^{\rm LC}$&$4.2_{-2.2}^{+2.9}$ &$3.5_{-1.5}^{+3.5}$ &$2.8_{-0.79}^{+4.2}$\\
      $b^{\rm VLC}$&$-20_{-17}^{+6.9}$&$-22_{-15}^{+8.8}$ &$-23_{-17}^{+9.7}$\\
      $\sigma^{\rm VLC}$&$0.14_{--0.13}^{+2.4}$&$0.59_{-0.35}^{+1.9}$ &$0.52_{-0.25}^{+2.1}$\\
      $b^{\rm HARPSpre}$&$-25_{-17}^{+7.5}$&$-27_{-15}^{+9}$&$-29_{-17}^{+9.9}$\\
      $\sigma^{\rm HARPSpre}$&$1.5_{-0.076}^{+0.055}$&$1.4_{-0.024}^{+0.11}$&$1.4_{-0.034}^{+0.099}$\\
      $w_1^{\rm HARPSpre}$&$0.67_{-0.027}^{+0.09}$&$0.68_{-0.038}^{+0.081}$&$0.68_{-0.038}^{+0.078}$\\
      $w_2^{\rm HARPSpre}$&$0.34_{-0.10}^{+0.028}$&$0.31_{-0.066}^{+0.067}$ &$0.35_{-0.11}^{+0.024}$\\
      ${\rm ln}{\frac{\tau^{\rm HARPSpre}}{\rm 1 day}}$&$0.97_{-0.18}^{+0.26}$&$1.1_{-0.34}^{+0.082}$&$1.0_{-0.21}^{+0.21}$\\
      $b^{\rm HARPSpost}$&$-12_{-18}^{+7.2}$&$-17_{-13}^{+11}$&$-15_{-18}^{+9.5}$ \\
      $\sigma^{\rm HARPSpost}$&$1.0_{-0.083}^{+0.21}$&$1.0_{-0.056}^{+0.24}$&$1.1_{-0.19}^{+0.11}$\\
      $w_1^{\rm HARPSpost}$& $0.92_{-0.051}^{+0.068}$& $0.97_{-0.097}^{+0.016}$&$0.92_{-0.053}^{+0.062}$ \\
      ${\rm ln}{\frac{\tau^{\rm HARPSpost}}{\rm 1 day}}$&$2.6_{-0.77}^{+3.9}$ & $6_{-4.2}^{+0.27}$&$2.7_{-0.9}^{+3.7}$\\
      $b^{\rm UVES}$&$-12_{-17}^{+7.6}$ &$-15_{-15}^{+9}$ &$-16_{-17}^{+9.7}$\\
      $\sigma^{\rm UVES}$&$0.63_{-0.14}^{+0.085}$& $0.6_{-0.11}^{+0.12}$ &$0.58_{-0.086}^{+0.15}$\\
      ln(BF)&314&314&314\\\hline
%      $P(e)=2\mathcal{N}(0,\,0.2)$ &$25_{-4.5}^{+13}$&$47_{-11}^{+13}$&$0.26_{-0.058}^{+0.061}$&$58_{-10}^{+30}$&$165_{-69}^{+40}$&314\\[2ex]
%       $P(e)=2\mathcal{N}(0,\,0.4)$ &$29_{-7.9}^{+9.3}$& $43.98_{-7.46}^{+17.66}$&$0.26_{-0.034}^{+0.090}$&$80_{-31}^{+12}$&$136_{-38}^{+70}$&314\\[2ex]
%      $P(e)=1$ &&&&$64_{-18}^{+32}$&&&&&&314\\
    \end{tabular}
    \label{tab:test}
  \end{table}

\section{Diagnostics of signals using Bayes factor periodograms} \label{sec:signal}
In this section we focus on the constraints provided by the long-term HARPS dataset alone and investigate evidence for other signals within the data sets of HARPS, HARPSlow, and HARPSpre using the {\small Agatha} software \citep{feng17a}, which is essentially a framework of red noise periodograms. By comparison of different noise models in Agatha, we find that a variety of optimal noise models: (1) HARPS --- fifth order moving average (or MA(5)) in combination with FWHM and NaD1, (2) HARPSlow --- MA(2) combined with BIS and NaD1, (3) HARPSpre --- MA(2) combined with S-index and NaD1.

\begin{figure}
  \begin{center}
% \hspace{-0.in}
\includegraphics[scale=0.4]{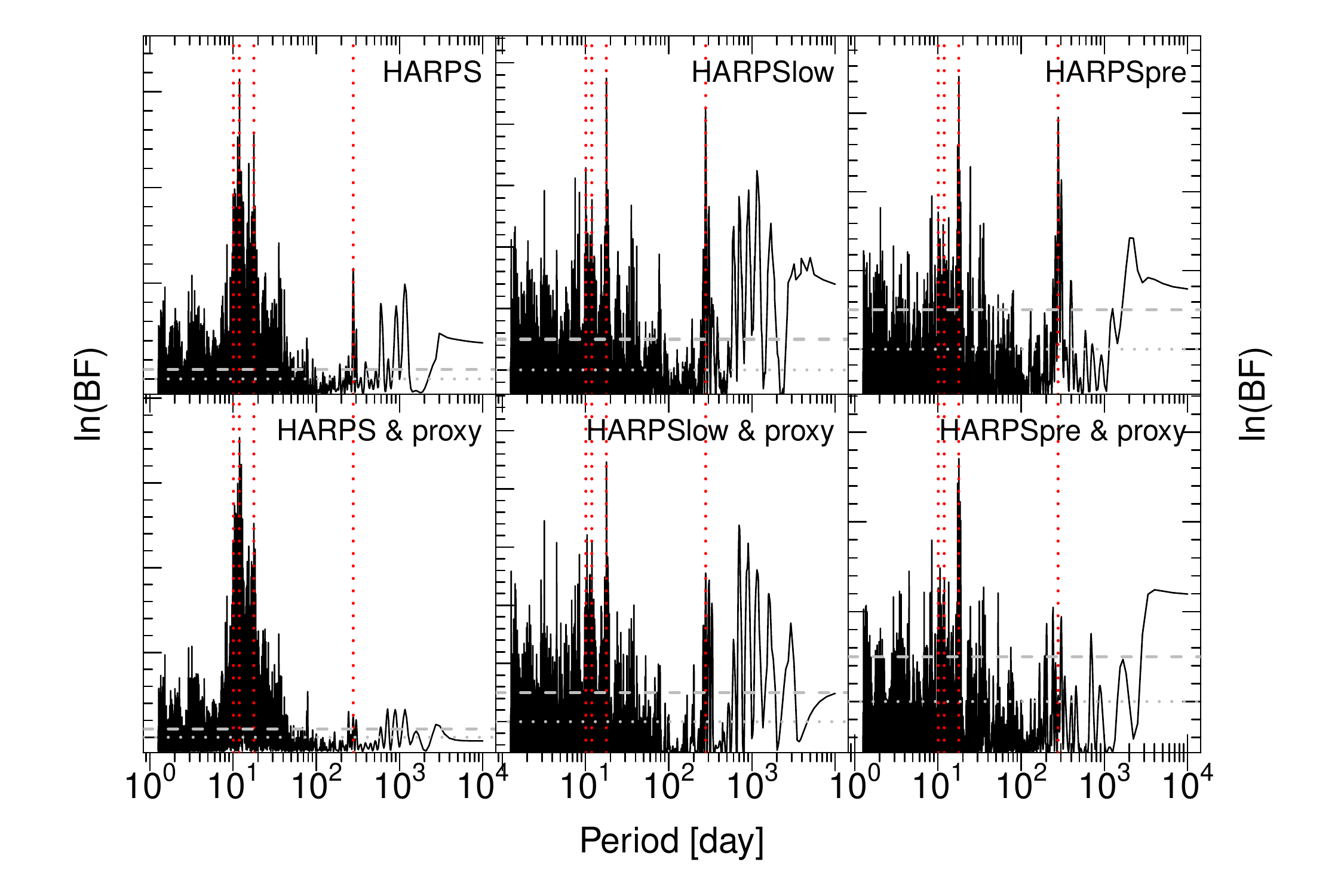}
  \caption{BFPs for MA(1) model in combination with noise proxies for the HARPS (left), HARPSlow (middle) and HARPSpre (right) data sets. The upper panels show the BFPs for the MA(1) model while the lowers ones for the MA(1) model in combination with the optimal noise proxies. The signals at periods of 10, 12, 18 and 278\,d are shown by dotted red lines. The thresholds of $\ln({\rm BF})=0$ and 5 are shown by the horizontal dotted and dashed lines, respectively. }
\label{fig:BFP1sig}
\end{center}
\end{figure}

To find primary signals, we calculate the Bayes factor periodogram (BFP) for the MA(1) model in combination with proxies for different data sets, and show them in Fig. \ref{fig:BFP1sig}. In this figure, the signal around 11\,d is significant in the HARPS data set because the high cadence data favors short period signals. There are also strong powers around this signal for the HARPSlow and HARPSpre data sets even though the high cadence and HARPSpost epochs have been excluded. For all data sets, the signal at 18\,d is significant although the HARPS data set favors the 11\,d signal more. The rather strong and noisy power around 11\,d in the left panels are contributed by 11 days of consecutive data which comprise the high cadence RVs.

In particular, the signal at a period of 278\,d is significant in the BFPs for MA(1) but become much weaker in the BFPs for MA(1) combined with noise proxies, suggesting an activity origin. The other strong signals in these BFPs are either aliases or harmonics of these three signals, as we will see in the subsequent analysis. 

\begin{figure*}
  \begin{center}
\includegraphics[scale=0.55]{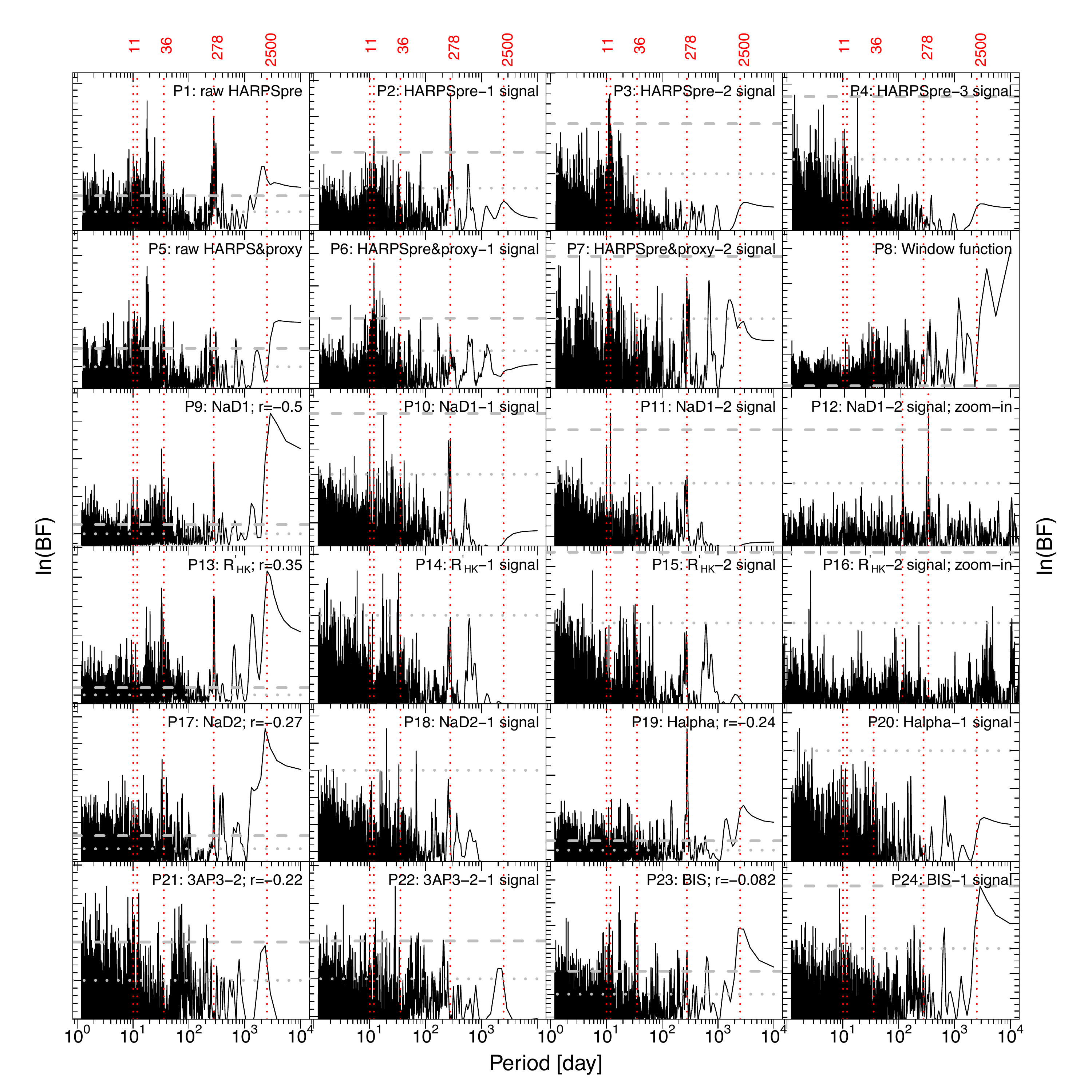}
  \caption{BFPs for RVs and noise proxies. Top row: BFPs for MA(1) for raw HARPSpre and  HARPSpre subsequently subtracted by signals at periods of 11, 18, 36, 278, and 2500\,d. The 11\,d signal is denoted by two periods at 10 and 12\,d according to posterior samplings. These periods are shown in the top of the figure. Second row: BFPs for MA(1) combined with S-index and NaD1 for raw HARPSpre and HARPSpre subtracted in turn by signals at periods at 18 and 11\,d, and the Lomb-Scargle periodogram for the window function (P8). The third row onward shows BFPs for MA(1) for noise proxies and their residuals. P12 and P16 are respectively the zoom-in versions of P11 and P15 and have a period range of $[5,~20]$\,d. For each panel, the Pearson correlation coefficient in the top right corner shows the correlation between noise proxies and RVs which are detrended by a second order polynomial. All panels are denoted by ``Pn'' with n varying from 1 to 24. The horizontal dotted and dashed lines denote $\ln{\rm BF}=0$ and 5, respectively. The BFP for the FWHM is not shown because there is not significant periodicity. The values of ln(BF) are indicated by thresholds 0 and 5 rather than shown by axis labels since different panels have different ranges of ln(BF). }
\label{fig:diagnostic}
\end{center}
\end{figure*}

Since the HARPSpre data set is more conservative than the HARPShigh and HARPSlow sets due to its sampling and is not subject to the uncertainty in radial velocity offset due to fibre change, we calculate the BFPs for the raw HARPSpre data and the RV residual signals and for various noise proxies, and show these BFPs in Fig. \ref{fig:diagnostic}. In panels P1-P14, we observed that the signals at periods of about 11, 18, and 278\,d are significant in the data. The 278\,d signal is likely caused by activity because its significance is reduced after accounting for the correlation between RVs and noise proxies, as seen in P5-P7. This signal is found to be significant in the BFPs of the NaD1, $R_{\rm HK}$, NaD2, and H$\alpha$ indices. In addition, the signal at a period of around 2500\,d is also significant in the BFPs of NaD1 (P9), R$'_{HK}$ (P13), NaD2 (P17), and BIS (P23 and P24). We interpret this signal as the magnetic cycle and the 278\,d signal as a secondary cycle in the magnetic variation. We also observe strong signals around 18\,d and its double period 36\,d in NaD1, NaD2, R$'_{HK}$, and BIS. Nevertheless, the 18\,d signal does not disappear in the BFPs accounting for linear correlations between RVs and NaD1 and R$'_{HK}$. Hence we conclude that this signal is either due to a nonlinear effect of stellar rotation or due to a planet, which has an orbital period similar to rotation period. The signal at about 11\,d is found to be unique and significant in the residual of NaD1 after subtraction of the 2500 and 18\,d signals. Thus this signal is probably caused by stellar activity or is an alias of activity-induced signals. We also look into the ASAS and Hipparcos photometric data but find few useful epochs and identify no significant signals. 

Therefore we conclude that the primary and secondary magnetic cycles
of $\epsilon$ Indi are 2500 and 278\,d. The rotation period of
$\epsilon$ Indi is about 36\,d, approximately double 18\,d. This rotation period is rather different from the 22\,d
value estimated by \cite{saar97} from Ca II measurements. Considering
that the 36\,d rotation period is derived from a relatively large
dataset of high precision RVs and multiple activity indicators, we
believe that 36\,d is a more reliable value of rotation period. On the
other hand, the half rotation period, 18\,d, is more significant
than the rotation period in the RV data. This phenomenon is also found
in the RVs of other stars (e.g. HD 147379; \citealt{feng18d}), and is
probably caused by a spot or spot complexity which more significantly
modulates spectral lines over one half of the rotation period. The
signal at a period of about 11\,d is also related to the stellar
rotation since $1/(1/36+1/18)\approx11$. Based on the Bayesian
quantification of these signals using the MA(1) model
\citep{tuomi12,feng16}, the semi-amplitudes of the signals at periods
of 10.064$^{+0.003}_{-0.001}$\footnote{Since the 10 and 12\,day
  signals are aliases of each other, we only report one signal
  here. }, 17.866$^{+0.006}_{-0.003}$ and 278$^{+1.7}_{-0.60}$\,d
signals are 1.4$_{-0.31}^{+0.18}$, 2.1$_{-0.19}^{+0.37}$, and
2.3$_{-0.31}^{+0.34}$\,m/s, respectively. The non-detection of
additional signals puts an upper limit of 1\,m/s on the semi-amplitude
RV variation induced by potential planets with periods less than the HARPSpre baseline (i.e. $P<4000$\,days) in the system. 

Based on the above analysis, different noise proxies are sensitive to different stellar activities. NaD1 and R$'_{HK}$ are sensitive to all activity-induced signals while BIS is sensitive to stellar rotation signal. H$\alpha$ is only sensitive to the secondary magnetic cycle. Considering the differential sensitivity of activity indicators to activity signals, a model including a linear correlation between RVs and these indicators would remove activity-noise as well as introduce extra noise. This is evident from the comparison of P1 with P5 in Fig. \ref{fig:diagnostic}. Although the 278\,d signal disappears after accounting for the linear correlation with S-index and NaD1, a longer period around 3000\,d and some short-period signals become significant (also compare P2 and P6, P3 and P7). Such extra signals are caused by fitting an inappropriate linear function of noise proxies to RVs while these proxies are not perfectly correlated with RVs linearly. Hence these proxies would bring their own signals into the likelihood/posterior distribution, which is a combined function of data and model, in order to mitigate the main activity signals in RVs. While the traditional activity indicators are limited in removing activity-induced signals, differential RVs are better in removing wavelength-dependent signals without introducing extra noise because they provide a simple way to weight spectral orders {\it a posteriori} \citep{feng17c}. Hence a reliable diagnostics of the nature of signals is to test the sensitivity of signals to different noise models, data chunks, and wavelengths. 

Since the activity-induced RV variations are not strictly periodic, Keplerian functions or a Gaussian process model cannot adequately model them. Hence the fit of a one-planet model is appropriate for parameter inference (see section \ref{sec:detection}). 

\section{Combined analysis of RV and astrometry}\label{sec:combined}
\subsection{Combined model of RV and astrometry}
In a star-planet or more generally star-companion system, the position
of the star in its orbital plane is calculated according to Kepler’s
equation is
\begin{equation}
  \bm{r}_s(t)=
  \begin{bmatrix}
    x(t)\\
    y(t)\\
    0
  \end{bmatrix}
  =\frac{m_p}{m_p+m_s}a
  \begin{bmatrix}
    \cos{E(t)}-e\\
    \sqrt{1-e^2}\sin{E(t)}\\
    0
  \end{bmatrix}~,
  \label{eqn:rs}
\end{equation}
where $m_s$ and $m_p$ are respectively the masses of star and planet,
$a$ is the semi-major axis of the planet with respect to the star,
$E(t)$ is the eccentricity anomaly, and $e$ is the eccentricity. The
corresponding stellar reflex velocity is
\begin{equation}
  \bm{v}_s(t)=
  \begin{bmatrix}
    v_x(t)\\
    v_y(t)\\
    0
  \end{bmatrix}
  =\frac{Gm_p^2}{(m_p+m_s)a(1-e^2)}
  \begin{bmatrix}
    -\sin{\nu(t)}\\
    \cos{\nu(t)}+e\\
    0
  \end{bmatrix}~,
    \label{eqn:vs}
\end{equation}
  where $\nu(t)$ is the true anomaly at time $t$, $G$ is the
  gravitational constant. The stellar position $\bm{r}_s(t)$ is
  converted to the observer frame coordinates, $\bm{r}_s^{\rm obs}
  (t)$, by applying Euler rotations using 
  \begin{equation}
    r_s^{\rm obs}=R_Z(\Omega)R_X(-I)R_Z(\omega)\bm{r}_s(t)~,
  \end{equation}
  where $\Omega$ is the longitude of ascending node, $\omega$ is the
  argument of periastron, $I$ is the inclination of the stellar orbit
  with respect to the sky plane. The directions of X axis (along North
  of the sky plane), Y axis (along East of the sky plane), and Z axis
  (along the line of sight towards the observer) forms a right-handed
  Cartesian coordinate system in the observer frame. The expansion of
  equation (5) gives the observed location of star in the XYZ
  coordinate system, 
  \begin{equation}
  \begin{bmatrix}
    x^{\rm obs}(t)\\
    y^{\rm obs}(t)\\
    z^{\rm obs}(t)\\
  \end{bmatrix}
  =
  \begin{bmatrix}
    A&F&-\sin{\Omega}\sin{I}\\
    B&G&\cos{\Omega}\sin{I}\\
    -\sin{\omega}\sin{I}&-\cos{\omega}\sin{I}&\cos{I}
  \end{bmatrix}
    \begin{bmatrix}
    x(t)\\
    y(t)\\
    0
  \end{bmatrix}
  ~,
\end{equation}
where
\begin{align}
  A&=\cos{\Omega}\cos{\omega}-\sin{\Omega}\sin{\omega}\cos{I}\\
  B&=\sin{\Omega}\cos{\omega}+\cos{\Omega}\sin{\omega}\cos{I}\\
  F&=-\cos{\Omega}\sin{\omega}-\sin{\Omega}\cos{\omega}\cos{I}\\
  G&=-\sin{\Omega}\sin{\omega}+\cos{\Omega}\cos{\omega}\cos{I}
\end{align}
are the so-called Thiele Innes constants. Hence the variation of right
ascension, declination, and parallax caused by a planet at a given
time is
\begin{align}
\alpha^p(t)&=\tilde{\omega}(t)y^{\rm obs}(t)=\tilde{\omega}(t)[Bx(t)+Gy(t)]~,\nonumber\\
\delta^p(t)&=\tilde{\omega}(t)x^{\rm obs}(t)=\tilde{\omega}(t)[Ax(t)+Fy(t)]~,
\label{eqn:adp}
\end{align}
The stellar reflex velocity is the time derivative of  $\bm{r}_s^{\rm
  obs} (t)$, which is
\begin{align}
\mu_\alpha(t)&=\tilde{\omega}(t)[Bv_x(t)+Gv_y(t)]~,\nonumber\\
\mu_\delta^p(t)&=\tilde{\omega}(t)[Av_x(t)+Fv_y(t)]~,
\label{eqn:muadp}
\end{align}
The planet-induced RV is
\begin{equation}
  v_r^p=-v_z^{\rm obs}=v\sin{I}[\cos{(\omega+\nu)}+e\cos{\omega}]~.
\label{eqn:vrp}
\end{equation}
The stellar motion is the sum of the barycentric motion (denoted by
superscript $b$) and the reflex motion (denoted by superscript $p$)
described by
\begin{align}
  \alpha(t)&=\frac{\alpha^p(t)}{\cos{\delta(t)}}+\alpha^b(t)~,\nonumber\\
  \delta(t)&=\delta^p(t)+\delta^b(t)~,\nonumber\\
  \frac{1}{\tilde{\omega}}&=-z^{\rm obs}(t)+\frac{1}{\tilde{\omega}^b(t)}~,\nonumber\\
  \mu_\alpha(t)&=\mu_\alpha^p(t)+\mu_\alpha^b(t)~,\nonumber\\
  \mu_\delta(t)&=\mu_\delta^p(t)+\mu_\delta^b(t)~,\nonumber\\
  v_r(t)&=v_r^p(t)+v_r^b(t)~.
\label{eqn:reflex}
\end{align}
Note that the distance or the inverse of parallax rather than the
parallax is additive. $\alpha^p(t)$ is a planet-induced offset in the sky plane and thus should be divided by $\cos{\delta(t)}$ to calculate the real change in righ ascension. Assuming zero acceleration, the heliocentric motion of the barycenter of a planetary system is determined by
\begin{align}
  \bm{v}^b(t)&=\bm{v}^b(t_0)\nonumber\\
  \bm{r}^b(t)&=\bm{r}^b(t_0)+\bm{v}^b(t_0)(t-t_0)
  \label{eqn:bary_motion}
\end{align}
where $\bm{v}^b(t)$ is the barycentric velocity, and $\bm{r}^b(t)$ is
the barycentric position. The linear motion is a good approximation of
heliocentric motion of the barycenter over
a few decades since the Galactic acceleration for nearby stars is typically less than mm/s/year.

Since the RV variation induced by the barycentric motion, $v_r^b (t)$,
is subtracted from the RV data, we only use $v_r^p (t)$ defined in
equation (9) to model the Keplerian signal in RV data. Considering the
barycentric RV is used to calculate the astrometry variation caused by
the change of perspective, the planet-induced RV variation or
gravitational redshift and convection blueshift only contribute as a
secondary variation in astrometric data. Thus we approximate the
barycentric RV by the Gaia RV if available or the value from the RV
data sets we collected, i.e. $v_r^b (t)\approx v_r (t)$.  We also
neglect the parallax variation induced by stellar reflex motion and by
the RV because they are secondary effects compared with the variation
of other observables. For example, the parallax of $\epsilon$ Indi A
is only changed by about 0.3\,$\mu$as due to the barycentric motion
along the line of sight. Thus we only model the astrometry observables
$\alpha$, $\delta$, $\mu_\alpha$ and $\mu_\delta$ of Hipparcos and
Gaia DR2 to constrain the planetary orbit. The procedure of astrometry
modeling is as follows.
\begin{itemize}
  \item Choose the reference time at the Gaia epoch J2015.5 and
    determine the initial barycentric astrometry according to 
    \begin{align}
      \alpha^b(t_0)&=\alpha^{\rm
                     gaia}(t_0)-\frac{\alpha^p(t_0)}{\cos{\delta^{\rm gaia}(t_0)}}-\Delta\alpha~,\\
      \delta^b(t_0)&=\delta^{\rm gaia}(t_0)-\delta^p(t_0)-\Delta\delta~,\\
      \tilde{\omega}(t_0)&=\tilde{\omega}(t_0)~,\\
      \mu_\alpha^b(t_0)&=\mu_\alpha^{\rm gaia}(t_0)-\mu_\alpha^{\rm p}(t_0)-\Delta\mu_\alpha~,\\
      \mu_\delta^b(t_0)&=\mu_\delta^{\rm gaia}(t_0)-\mu_\delta^{\rm p}(t_0)-\Delta\mu_\delta~,\\
      v_r^b(t_0)&=v_r(t_0)~,
    \end{align}
    where $\Delta\alpha$, $\Delta\delta$, $\Delta\mu_\alpha$, and
    $\Delta\mu_\delta$ are the offsets. These offsets are used to
    account for the reference frame spin rate (about 0.15\,mas/yr
    according to \cite{lindegren18}), for the zero point offset of
    parallaxes (about -0.029\,mas according to \cite{lindegren18}),
    and for the proper motion offsets related to light travel time (about 0.6 mas/yr in the case of $\epsilon$ Indi A; \citealt{kervella19}), and for other effects. 
  \item Transform initial astrometry and RV into heliocentric state
    vector $[\bm{r}^b(t_0),\bm{v}^b(t_0)]$, and propagate the vector
    from $t_0$ to $t$ to derive$[\bm{r}^b(t),\bm{v}^b(t)]$ according
    to equation \ref{eqn:bary_motion}, and transform the state vector
    back to barycentric astrometry at time t. This step is to
    calculate perspective acceleration.

  \item Estimate $\alpha(t)$, $\delta(t)$, $\mu_\alpha(t)$, and
    $\mu_\delta (t)$ by summing the barycentric and reflex astrometry
    according to equations \ref{eqn:adp}, \ref{eqn:muadp}, and \ref{eqn:reflex}. 
  \end{itemize}
To account for unknown noise in astrometry data, we include a relative
jitter in units of observational errors in the likelihood. Thus the
likelihood of the combined RV and astrometry model is
\begin{align}
  \ln\mathcal{L}&\equiv P(D|\theta,M)=\prod_k^{N_{\rm
      set}}\prod_i^{N_k^{\rm
      rv}}\frac{1}{\sqrt{2\pi(\sigma_j^2+\sigma_{\rm
        Jk}^2)}}\exp\left\{-\frac{[\hat{v}_r(t_i)-v_r(t_i)^2]}{2(\sigma_i^2+\sigma_{\rm Jk}^2)}\right\}\nonumber\\
  &+(2\pi)^{\frac{N_{\rm epoch}N_{\rm par}}{2}}\prod_j^{N_{\rm epoch}}({\rm det}\Sigma_j)^{-\frac{1}{2}}{\rm exp}\left\{-\frac{1}{2}[\hat{\bm{\eta}}(t_i)-\bm{\eta}(t_i)]^T\Sigma_j^{-1}[\hat{\bm{\eta}}(t_i)-\bm{\eta}(t_i)]\right\}~,
\label{eqn:comblike}
\end{align}
where $N_{\rm set}$, $N_{\rm epoch}$, and $N_{\rm par}$  are
respectively the number of RV data sets, astrometry epochs, and free
parameters of the astrometry model. $N_k^{\rm rv}$ is the number of
RVs in the $k^{\rm th}$ RV set, $\bm{\eta}\equiv
[\alpha,\delta,\mu_\alpha,\mu_\delta]$ is the astrometry data, and
$\Sigma$ is the jitter corrected covariance matrix of $\bm{\eta}$,
$\Sigma_j\equiv \Sigma_{\rm 0j}(1+J_j)$ where $\Sigma_{\rm 0j}$ is the
catalog covariance matrix for the $j^{\rm th}$ astrometry epoch, and
$J_j$ is the so-called ``relative astrometry jitter''. These relative
jitters allow the model to account for potential underestimation of
astrometry errors such as the so-called DOF bug \citep{lindegren18}
and the effects mentioned in \cite{brandt18}. Unlike previous
calibrations of the five-parameter solutions of Hipparcos and Gaia DR2
based on statistical analyses \citep{lindegren18,brandt18}, we model
the potential underestimation of uncertainties and bias as offsets and
astrometric jitters, and estimate them a posteriori rather than a
priori by taking advantage of the high precision RV data and a
combined modeling of the barycentric and reflex motions of $\epsilon$ Indi A.
The RV model $\hat{v}_r(t_i)$ is given in Eqn. \ref{eqn:rv1}.

Therefore the free parameters in the combined model are
$\{m_{pk},a_k,e_k,I_k,\omega_k,\Omega_k,M_{0k}\}$ with
$k\in\{1,...,N_p\}$ for $N_p$ planets or companions,
$\{\Delta\alpha,\Delta\delta,\Delta\mu_\alpha,\Delta\mu_\delta,J_{\rm
  gaia},J_{\rm hip} \}$ for astrometry,$\{b_1,...,b_{N_{\rm set}}\}$
for the offsets and $\sigma_1,...,\sigma_{N_{\rm set}}$ for the
jitters in $N_{\rm set}$ independent RV sets, $\{w_1^i,...,w_q^i\}$
and $\tau^i$ for the MA(q) model for the $i^{\rm th}$ RV data set. The
superscript of a given parameter denotes the name of the corresponding data set. The argument of perihelion and semi-major axis for a planetary orbit with respect
to the barycenter are $\omega_p=\omega+\pi$ and $a_p=m_s/(m_p+m_s) a$
while the other angular parameters are the same as for the stellar
reflex motion.

The offsets between Gaia and Hipparco catalog data are also used by
\cite{calissendorff18,brandt18,kervella19} to constrain the dynamic
mass of massive companions and to identify potential companion-induced
accelerations. In particular, \cite{snellen18} used both the proper
motion difference and Hipparcos epoch data to constrain the mass of
$\beta$ Pictoris b. While these studies justify use of astrometry
difference between Hipparcos and Gaia DR2 to constrain stellar reflex
motion, the constraints are not strong due to a lack of combined modeling of RV
and the difference in proper motion as well as in position. In
particular, we find that position difference is found to be more
sensitive to non-linear stellar motions which appear from
orbital integration over decades. The aim of this work is to focus on
a target with clear-cut RV and astrometric evidence alongside
appropriate treatment of Gaia and Hipparcos catalog data in
combination with RV analysis.

\subsection{Orbital solution}\label{sec:solution}
\begin{figure}
  \centering
\includegraphics[scale=1]{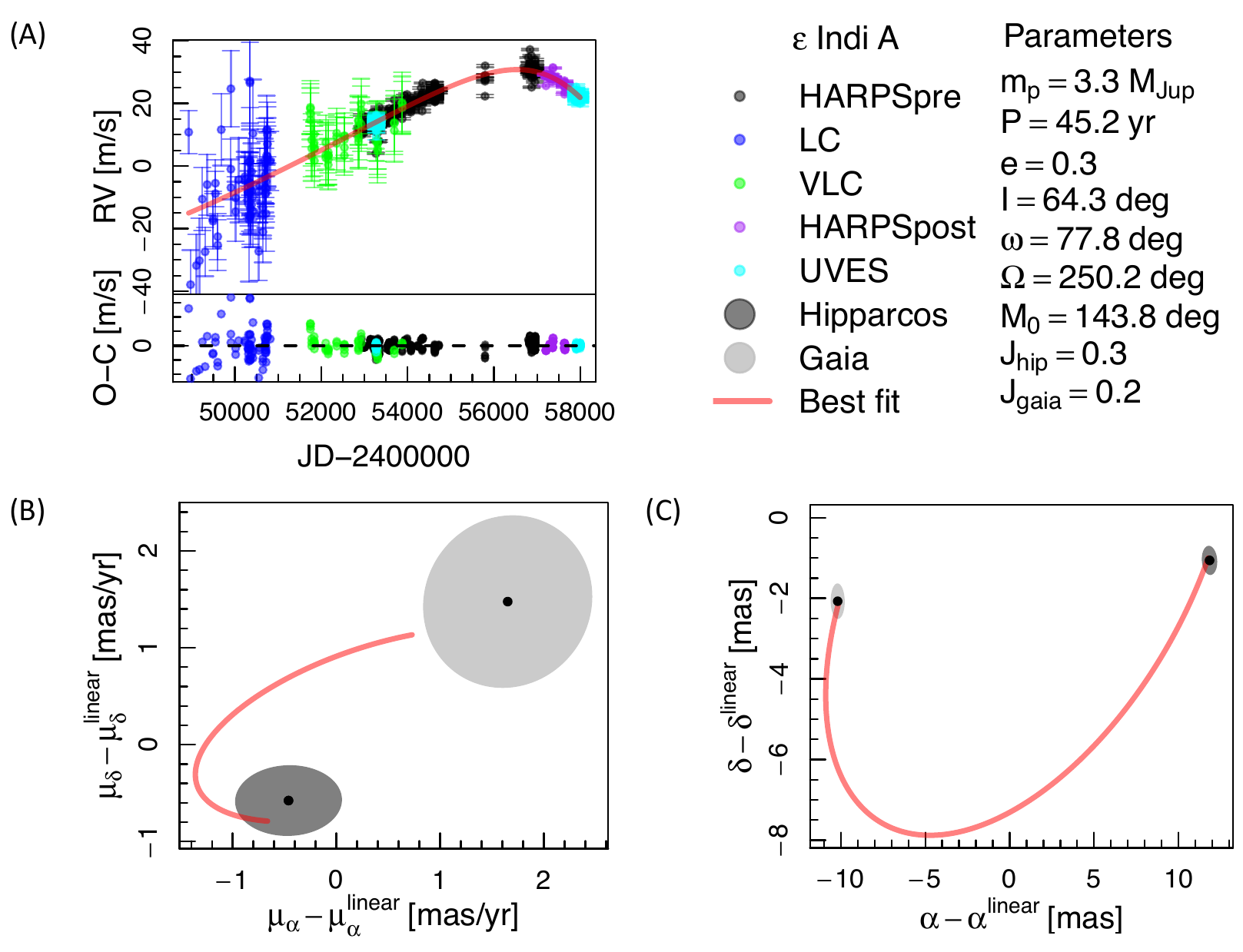}
\caption{Observed RV data (A) and astrometric data in terms of
  proper motion (B) and position (C) compared to best fit model
  predictions. The offsets due to linear motion are subtracted from
  the data. (A) RV data of $\epsilon$ Indi A compared to the best fit
  model parameters (maximum a posteriori) represented as a red line
  with values shown in the legend. $J_{\rm gaia}$ and $J_{\rm hip}$
  are respectively the relative excess noise in Gaia and Hipparcos
  astrometry data. (B) Hipparcos and Gaia DR2 proper motions are
  respectively denoted by the dark and light gray ellipse with a black
  dot in the center. The red line denotes the same ``best'' model
  prediction of the proper motion shown in (A). The ellipses show the
  90\% jitter-corrected confidence level of astrometry. (C) The
  relative Hipparcos position at BJD2448349.0625 and Gaia DR2 position
    at BJD2457206.375 are shown along with the red line showing the best model prediction of the position. The barycentric motion of $\epsilon$ Indi A is subtracted from data in order to show its reflex motion. }
\label{fig:fit2}
\end{figure}

Based on the MCMC sampling of the posterior, we find the best orbital
solution and show the fit to RV and astrometry data in
Fig. \ref{fig:fit2}. In panel (A), it is visually apparent that the
one-planet model is strongly favored relative to the zero-planet model
($\ln{\mathcal{L}}=331$ or ln\,BF=311) and that the curvature in
the RV data is significant compared with a linear trend
($\ln{\mathcal{L}}=51$ or ln\,BF=38). Fig. \ref{fig:fit2} (B) and (C) indicate that the position puts a stronger constraint on the best fit model of the planetary orbit than the proper motion and comparable to the highest precision RV data. This arises because the position difference serves to integrate the non-linear proper motion over time whereas the proper motion difference is not as sensitive as position data to non-linear motion due to potential bias caused by the assumption of linear motion in the production of catalog data.

Based on the combined analysis, we report the values of the orbital
parameters of $\epsilon$ Indi A b together with the stellar parameters
in Table \ref{tab:combined}. Note that the stellar mass slightly differs from
the value given by \cite{demory09} because we determine the mass using the Gaia luminosity and the mass-luminosity relationship derived by \cite{eker15}. The posterior distribution of these
parameters is shown in Fig. \ref{fig:pair}. In particular, we show
the posterior distribution of planetary mass and orbital period in
Fig. \ref{fig:mp}. It is apparent that the planetary mass
and orbital period are constrained to a relatively high precision even
without RV and astrometry data that cover the whole orbital phase.
  \begin{table}
    \centering
    \caption{Parameters for $\epsilon$ Indi A and A b are taken from
      Gaia DR2 \citep{brown18} or determined in this work. Except for
      rotation period and magnetic cycle for parameters determined in
      this work, the optimal value is estimated at the maximum a
      posteriori and the uncertainty interval is determined by the
      10\% and 90\% quantiles of the posterior samples drawn by the
      MCMC chains.}
    \renewcommand{\arraystretch}{1.2}
    \begin{tabular}{l*3{l}}
      \hline
      \hline
      Parameter&Unit&Meaning&Value\\\hline
      $m_s$&$M_\odot$&Stellar mass & $0.754_{-0.043}^{+0.043}$\\
      $L_s$&$L_\odot$&Stellar Luminosity& $0.239_{-0.001}^{+0.001}$\\
      $\alpha$&degree&Right ascension& 330.87\\
      $\delta$&degree&Declination& -56.80\\
      $\tilde{\omega}$&mas&Parallax& $274.8_{-0.25}^{+0.25}$\\  
      $\mu_\alpha$&mas/yr&Proper motion in right ascension& $3967.04_{-0.38}^{+0.38}$\\
      $\mu_\delta$&mas/yr&Proper motion in right ascension&$-2535.76_{-0.41}^{+0.41}$ \\
      $v_r$&km/s&Systematic radial velocity&$-40.50_{-0.23}^{+0.23}$
    \\
      $P_{\rm rot}$&day&Stellar rotation
                         period&$-35.732_{-0.003}^{+0.006}$ \\
      $P_{\rm mag}$&day&Stellar magnetic cycle&2500$\sim$3000 \\
      $T_{\rm age}$&Gyr&Stellar age&3.7$\sim$5.7\\\hline
      $m_p$&$M_{\rm Jup}$&Planet mass&$3.25_{-0.65}^{+0.39}$\\
      $P$&year&Orbital period&$45.20_{-4.77}^{+5.74}$\\
      $a$&au&Semi-major axis&$11.55_{-0.86}^{+0.98}$\\
      $K$&m/s&RV semi-amplitude&$29.22_{-6.07}^{+5.45}$\\
      $e$&&Eccentricity&$0.26_{-0.03}^{+0.07}$\\
      $I$&degree&Inclination&$64.25_{-6.09}^{+13.80}$\\
      $\Omega$&degree&Longitude of ascending node&$250.20_{-14.84}^{+14.72}$\\
      $\omega$&degree&Argument of periastron&$77.83_{-31.51}^{+20.21}$\\
      $M_0$&degree&Mean anomaly at reference
                    epoch&$143.8_{-58.75}^{+23.38}$\\
      $T_0$&BJD&Reference epoch&2448929.56\\
      $T_p$&BJD&Epoch at periastron&$2442332.95_{-3450.17}^{+2353.42}$\\\hline      
      $b^{\rm LC}$&m/s&LC offset& $-16.60_{-5.10}^{+4.33}$\\
      $b^{\rm VLC}$&m/s&VLC offset& $-22.97_{-3.84}^{+5.33}$\\
      $b^{\rm HARPSpre}$&m/s&HARPSpre offset&
                    $-27.87_{-4.08}^{+5.16}$\\
      $b^{\rm HARPSpost}$&m/s&HARPSpost offset&
                      $-14.23_{-5.75}^{+4.07}$\\
      $b^{\rm UVES}$&m/s&UVES offset& $-15.04_{-4.09}^{+5.03}$\\
      $\sigma^{\rm LC}$&m/s&LC RV jitter& $4.05_{-1.97}^{+3.01}$\\
      $\sigma^{\rm VLC}$&m/s&VLC RV jitter& $0.42_{-0.15}^{+2.11}$\\
      $\sigma^{\rm HARPSpre}$&m/s&HARPSpre RV jitter& $1.45_{-0.05}^{+0.09}$\\
      $\sigma^{\rm HARPSpost}$&m/s&HARPSpost RV jitter& $1.02_{-0.07}^{+0.22}$\\
      $\sigma^{\rm UVES}$&m/s&UVES RV jitter& $0.64_{-0.14}^{+0.09}$\\
      $w_1^{\rm HARPSpre}$&m/s&Amplitude of MA(2) for HARPSpre&$0.69_{-0.04}^{+0.07}$\\
      $w_2^{\rm HARPSpre}$&m/s&Amplitude of MA(2) for HARPSpre&$0.31_{-0.07}^{+0.06}$\\
      $\ln{\frac{\tau^{\rm HARPSpre}}{1~{\rm day}}}$&&Logarithmic MA
                    time
                                                       scale&$0.98_{-0.18}^{+0.24}$\\
            $w_1^{\rm HARPSpost}$&m/s&Amplitude of MA(1) for
     HARPSpost&$0.98_{-0.11}^{+0.00}$\\
            $\ln{\frac{\tau^{\rm HARPSpost}}{1~{\rm day}}}$&&Logarithmic MA
                    time
                                                              scale&$2.16_{-0.41}^{+4.22}$\\
      $J_{\rm hip}$&&Hipparcos relative jitter&$0.26_{-0.22}^{+0.08}$\\
      $J_{\rm gaia}$&&Gaia relative jitter&$0.19_{-0.14}^{+0.21}$\\
      $\Delta\alpha$&mas&Offset in $\alpha$ &$-0.99_{-0.09}^{+0.40}$\\
      $\Delta\delta$&mas&Offset in $\delta$ &$-0.01_{-0.26}^{+0.25}$\\
      $\Delta\mu_\alpha$&mas&Offset in $\mu_\alpha$ &$0.92_{-0.20}^{+0.29}$\\
      $\Delta\mu_\delta$&mas&Offset in $\mu_\delta$ &$0.34_{-0.30}^{+0.54}$\\
      \hline
    \end{tabular}
    \renewcommand{\arraystretch}{1}
    \label{tab:combined}
  \end{table}

\begin{figure*}
  \begin{center}
\includegraphics[scale=0.6]{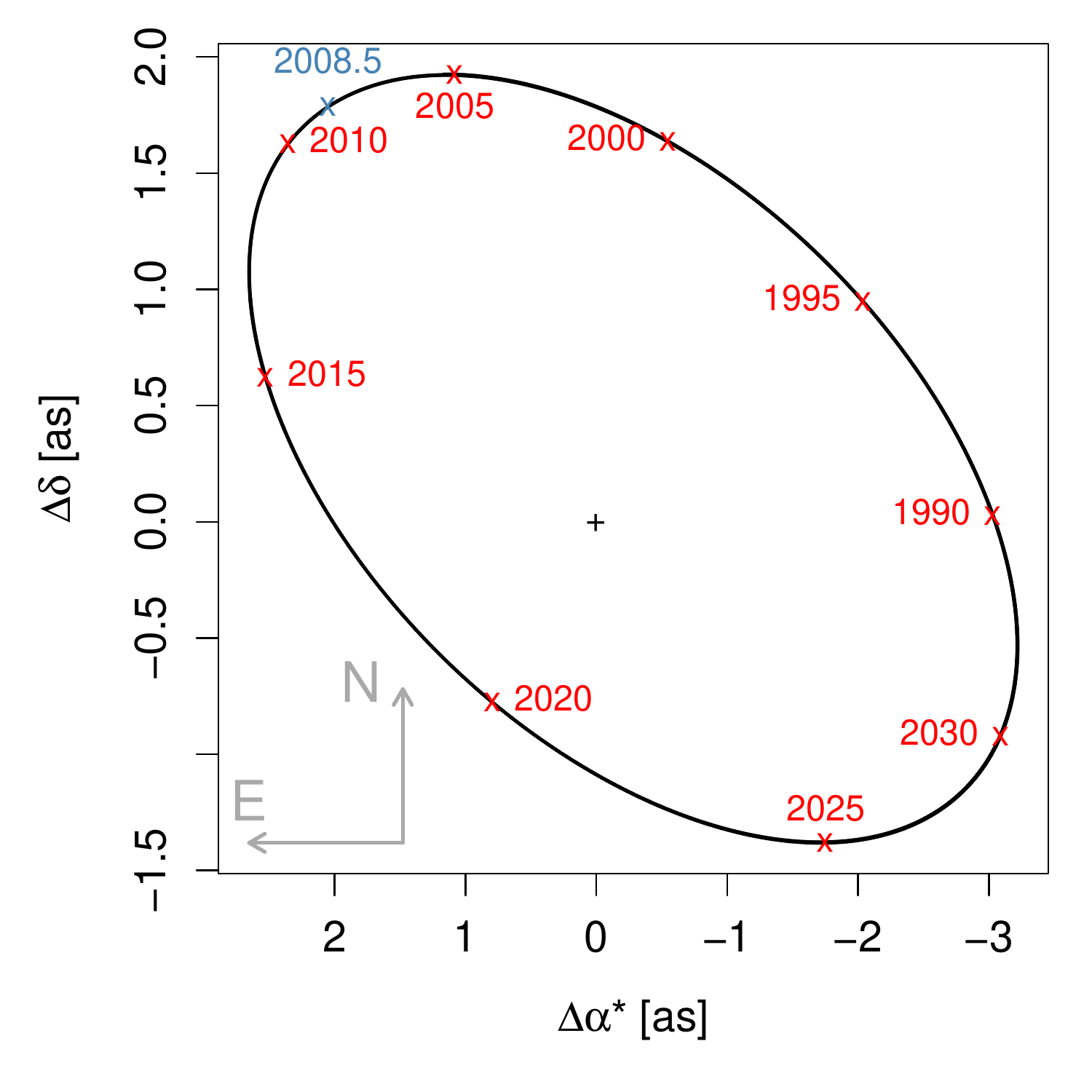}
\caption{Orbit of $\epsilon$ Indi A b relative to its host star
  projected onto the sky plane. The position at different epochs are
  given as references for direct imaging. The separation
  between the planet and the star is about 1.1\,as in 2020. The light
  blue cross indicates the average epoch of the direc imaging data used by
  \protect\cite{janson09}. The North (N) and East (E) directions are shown by
  grey arrows. The black cross indicates the location of the star. Note that the orbit is
  shown in offset coordinates and $\Delta\alpha*\equiv\Delta\alpha\cos{\delta}$. }
\label{fig:ei}
\end{center}
\end{figure*}

Based on the above analysis, $\epsilon$ Indi Ab is a cold Jovian
exoplanet with a very long orbital period among the exoplanets
detected through the radial velocity and transit methods.  With
components A, Ab, Ba and Bb, $\epsilon$ Indi provides a benchmark
system for the formation of gas giants and brown dwarfs. $\epsilon$
Indi Ab is also a perfect target for direct imaging. The non-detection
in previous direct imaging suggests a very low temperature
\citep{janson09}. Based on the new combined constraint in this work,
the orbit of $\epsilon$ Indi A b with respect to $\epsilon$ Indi A is shown in Fig. \ref{fig:ei}.
It is apparent that the current separation of the planet from the star
is optimal for direct imaging. The separation will increase from
1.1\,as in 2020 to 3.3\,as in 2030 and thus provides a
perfect chance for imaging of a nearby cold Jupiter by JWST
\citep{krist07} or WFIRST \citep{melchior17}. Althoug the separation
between $\epsilon$ Indi Ab and its host star at the epochs of previous
direct imaging \citep{janson09} is wider than the one in 2020,
our new constraint of the orbital parameters and mass of the planet
can be used to optimize the observation strategy. Moreover, $\epsilon$
Indi A b would change its host star's  position by about 1\,mas during Gaia's five year mission and thus is detectable by Gaia with its $\sim$20\,$\mu$as astrometric precision \citep{perryman14}. 

\subsection{Dynamical stability of $\epsilon$ Indi A b}
With one Jovian planet and a binary brown dwarf system, the $\epsilon$ Indi
system provides a benchmark case to test theories on the formation of
giant planets and brown dwarfs. The age of the system is about
0.8$\sim$2.0 Gyr based on the estimated literature rotation period of
22\,day determined from Ca II measurements. The kinematics of the
system show an older age of >9.87\,Gyr \citep{eker15}. Our analysis of the
HARPSpre data set gives a rotation period of 35.7\,d leading to an age
of 3.7$\sim$5.7 Gyr based on the rotation-age relationship given by
\cite{eker15} which is more consistent with the age of 3.7$\sim$4.3 Gyr
estimated by \cite{scholz03} based on an evolutionary model of the brown dwarf
binary Ba and Bb. Thus we conclude that the age of the system is about
4\,Gyr.

However, stars less massive than the Sun are unlikely to form more
than one giant planet according to \cite{kennedy07}. To investigate
whether B a and B b were captured by $\epsilon$ Indi A either in or out
of its birth environment, we calculate the escape radius derived by \cite{feng18} based on simulations of perturbations from stellar encounters on
wide binaries and an encounter rate of 80\,Myr$^{-1}$ for stellar encounters with
periapsis less than 1 pc. The escape radius of $\epsilon$ Indi A is about 5600
au which is larger than the projected separation (1459\,au) between
$\epsilon$ Indi B and A \citep{scholz03}. However, if $\epsilon$ Indi
A migrated outward to its current location or captured $\epsilon$ Indi
B during its formation, the encounter rate would be much higher and the escaped
radius could be within the orbit of $\epsilon$ Indi B, which may also
be much larger than the projected separation due to geometric
effects. On the other hand, the brown dwarf binary may have been on a
tighter orbit around the primary during the early evolution of the
system and have migrated to its current orbit due to perturbations
from the Galactic tide and stellar encounters.

Considering these scenarios of dynamical history and that the components of the $\epsilon$ Indi system are likely to have the same age, we conclude that the system was probably formed together and the brown dwarf binary has migrated from a tighter orbit to its current wide orbit. Such a migration would significantly influence the habitability of the system through periodic perturbations from the binary on any potentially habitable planets in the system. 

\subsection{Comparing $\epsilon$ Indi A b with known exoplanets}\label{sec:comparing}
\begin{figure*}
  \begin{center}
\includegraphics[scale=1]{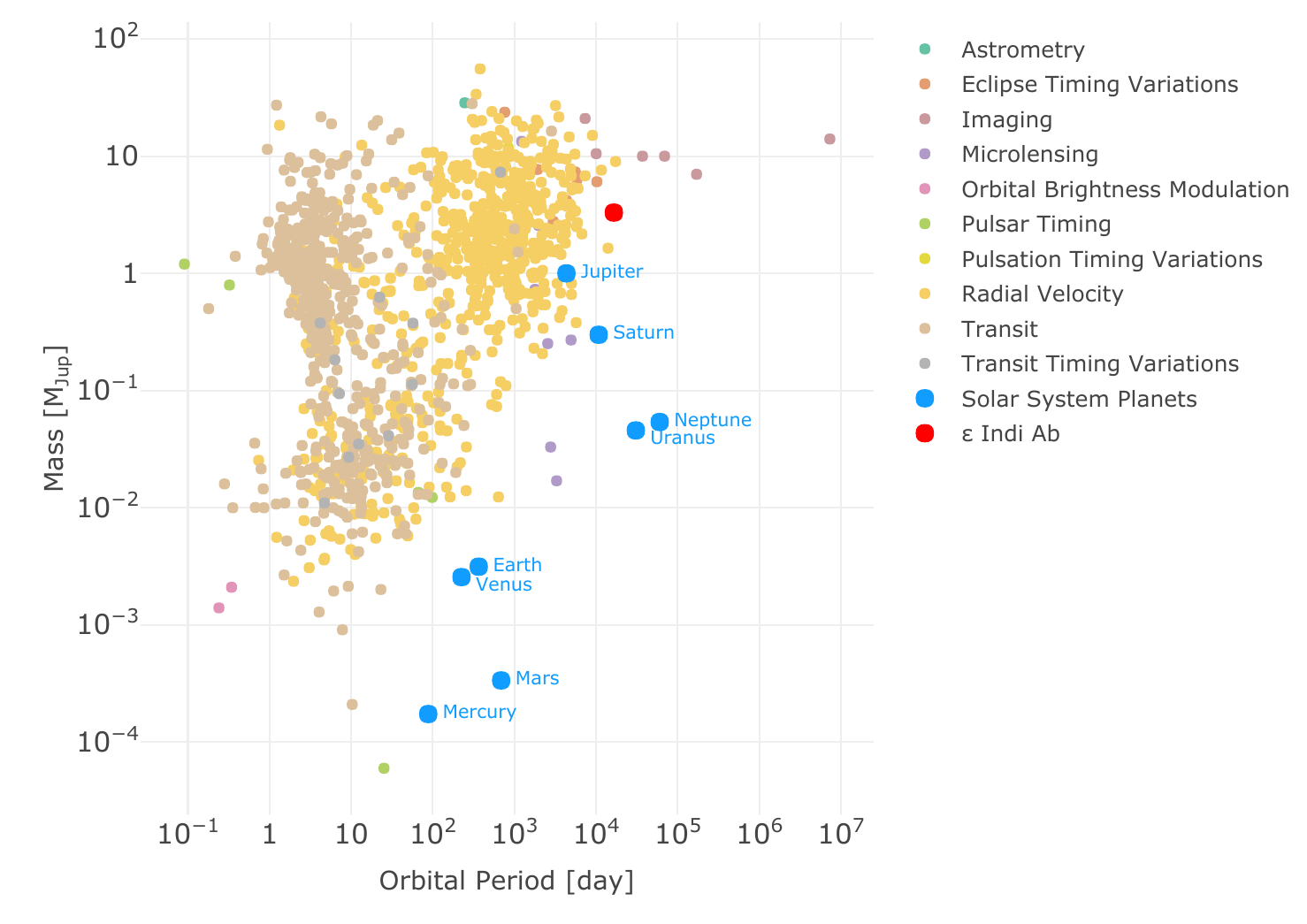}
\caption{Scatter plot of orbital period and planetary mass for confirmed exoplanets, Solar System planets, and $\epsilon$ Indi A b. The Solar System planets are denoted by their names. The exoplanets are downloaded from NASA Exoplanet Archive and are color-coded for different detection methods.}
\label{fig:epsb}
\end{center}
\end{figure*}

The discovery of the nearby Jupiter analog $\epsilon$ Indi A b is a
milestone for the studies of the formation and evolution of Jupiter
analogs. In Fig. \ref{fig:epsb}, we show the mass and period of $\epsilon$ Indi A b compared
with known exoplanets and the Solar System planets. We see that a few
Jupiter-like planets (with 0.5$\sim$5 Jupiter mass) have been detected
previously by the RV method though in these cases only provide minimum
masses. So far none of them are confirmed or characterized by other
methods. Although direct imaging is able to detect super-Jupiters and
brown dwarfs around young stars, it is not able to probe the region
below 8 Jupiter mass especially for evolved systems where Jupiter-like
planets become very faint. Given the limitations of various methods in
detecting wide-orbit gas giants, the detection of $\epsilon$ Indi A b
by two methods illustrates how a good constraint (e.g., Fig. \ref{fig:mp}) can
be set on a wide-orbit cold gas giant in an unpopulated region of
exoplanet phase space by using astrometry and RV together. 

\section{Discussions and conclusion}\label{sec:conclusion} 
We analyze the RV and astrometry data of $\epsilon$ Indi A in the
{\small Agatha} framework in combination with Bayesian methods. We confirm the
suspected planet $\epsilon$ Indi A b to be a cold Jovian planet on a
11.55\,au-wide orbit with an orbital period of 45.20\,yr, making it
the planet with a very long period from exoplanets detected by the RV
methods. $\epsilon$ Indi A b is only 3.62\,pc away from the Sun and is the closest Jovian exoplanet from the Earth. Given its
proximity to the Sun, $\epsilon$ Indi A b is separated from its host
star by as much as 3.5\,$''$ and thus can be observed through direct
imaging and astrometry, for example by JWST and Gaia
\citep{krist07,perryman14} though based on the \cite{baraffe03}
isochrones it is likely to be too faint for the current ground-based
imaging systems (e.g., \citealt{janson09}).

We also diagnose the other signals by comparing the BFPs for various
RV data sets, noise models, and noise proxies. We find three signals
at periods of about 11, 18 and 278\,d in the RV data of $\epsilon$ Indi A. These signals are significant and can be constrained by Bayesian posterior samplings. Nevertheless, they are unlikely to be Keplerian because significant powers around these signals are found in the periodograms of various noise proxies especially in sodium lines. Based on the activity diagnostics, we conclude that all these signals together with the 2500\,d signal in many noise proxies are caused by stellar activity. In particular, the 2500 and 278\,d signals correspond to the magnetic cycles of $\epsilon$ Indi A while the 35, 18 and 11\,d signals are related to the stellar rotation. We find that the correlation between RVs and activity indicators depends on activity time scales. Thus the inclusion of them in the fit would remove activity-induced noise as well as introduce activity-induced signals. 

The lack of signals with $K>1$\,m/s and $P<4000$\,days suggests a lack of super-Earths or
mini-Neptunes in the habitable zone of $\epsilon$ Indi A which is 0.47 to 0.86
au according to \citep{kopparapu14}. Hence only Earth-like or smaller
planets are allowed in the habitable zone. If these rocky planets are detected,
$\epsilon$ Indi A would be similar to the Solar System, with close-in
rocky planets and longer period gas giants. Thus, we also investigate
the dynamical stability of this system according to the metrics of the
stability of wide binaries under the perturbation of stellar
encounters and Galactic tide \citep{feng18}. We find a considerable possibility
that the brown dwarf binary in this system, $\epsilon$ Indi B a and B b, is
unstable if they are on their current wide orbit. Hence the brown
dwarf binary may have migrated from a tighter orbit to their current
wide orbit, which might significantly influence the habitability of
potentially Earth-like planet in the system.

Our successful detection and characterization of $\epsilon$ Indi A b
through combined RV and astrometry analysis provides a benchmark
example for the use of the astrometric difference between Gaia and
Hipparcos data in characterizing massive planets. We find that the
position offsets are more sensitive to nonlinear motion than the
proper motion offsets. Although there are only two astrometry epochs,
each epoch corresponds to four independent data points. Thus we are able to constrain the orbital parameters that the RV data is not
sensitive to. We eagerly anticipate joint application of astrometry
and RV methods in exoplanet detection and characterization with the
epoch data in future Gaia data releases. $\epsilon$ Indi A b is
optimal for direct imaging with a current spearation from its host of
about 1\,as. The separation will increase to 3.3\,as in the coming
decade, making it the nearest Jupiter analog for direct imaging by JWST or WFIRST.
  
\section*{Acknowledgements}
We used the ESO Science Archive Facility to collect radial velocity data. We are very grateful to the referee for comments which significantly clarified the presentation of the results and the quality of the manuscript.

\appendix
\section{Posterior distribution of model parameters}
\begin{figure*}
  \begin{center}
    \includegraphics[scale=0.45]{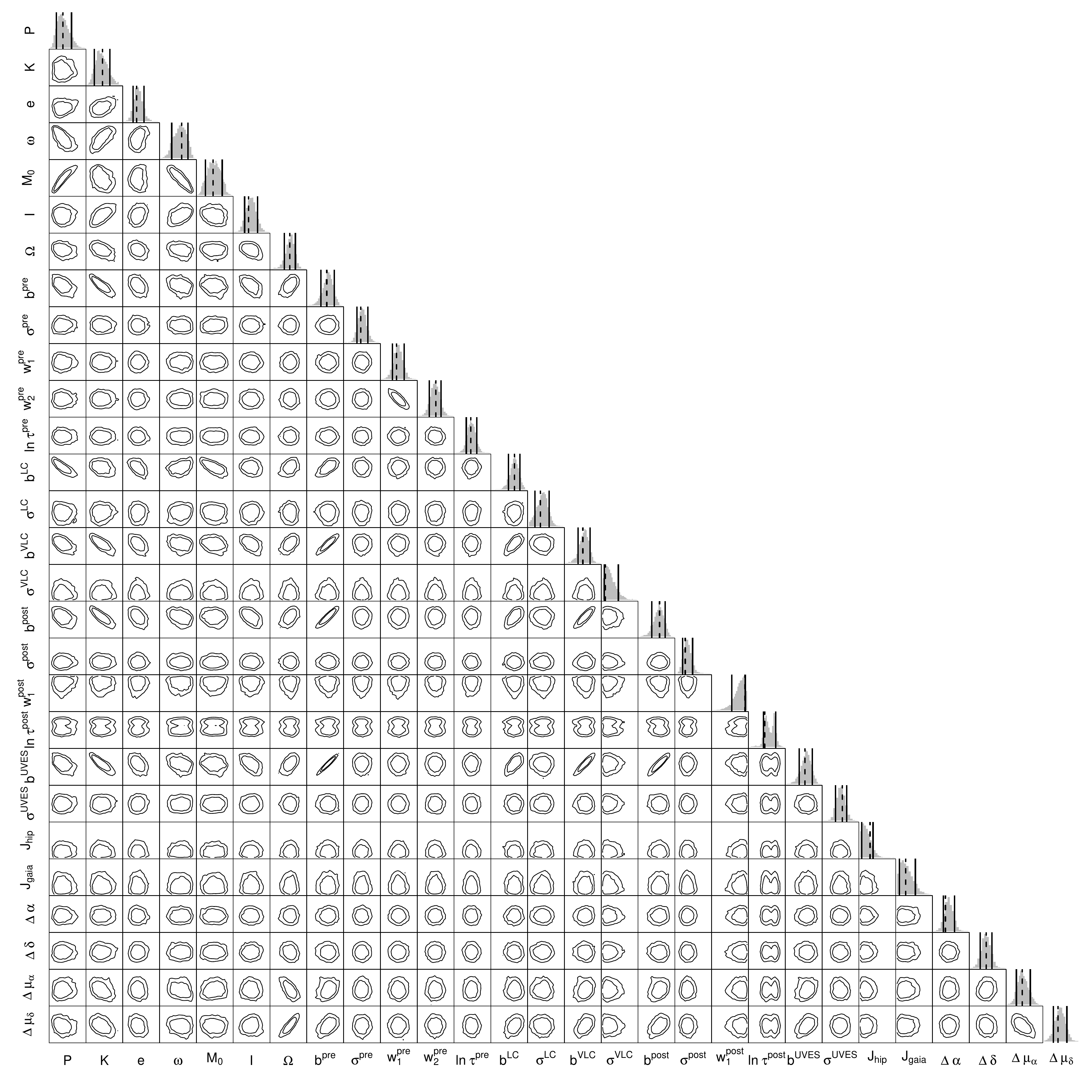}
\caption{Two-dimensional posterior distributions for all model
  parameters. The first five parameters are for the Keplerian signal,
  $b $ is offset, $\omega$ and $\ln{\tau}$ are respectively the amplitude and logarithmic time scale of the MA model. The names of data sets are denoted by the superscripts for relevant parameters. The ``pre'' and ``post'' superscripts denote the HARPSpre and HARPSpost sets, respectively. The contours show the 68\% and 90\% confidence levels. The dashed lines denote the MAP parameter values in the posterior distribution while the solid lines show the 10\% and 90\% quantiles of the distribution.}
\label{fig:pair}
\end{center}
\end{figure*}

\section{Posterior distribution of mass and period}
\begin{figure*}
\centering
\includegraphics[scale=0.8]{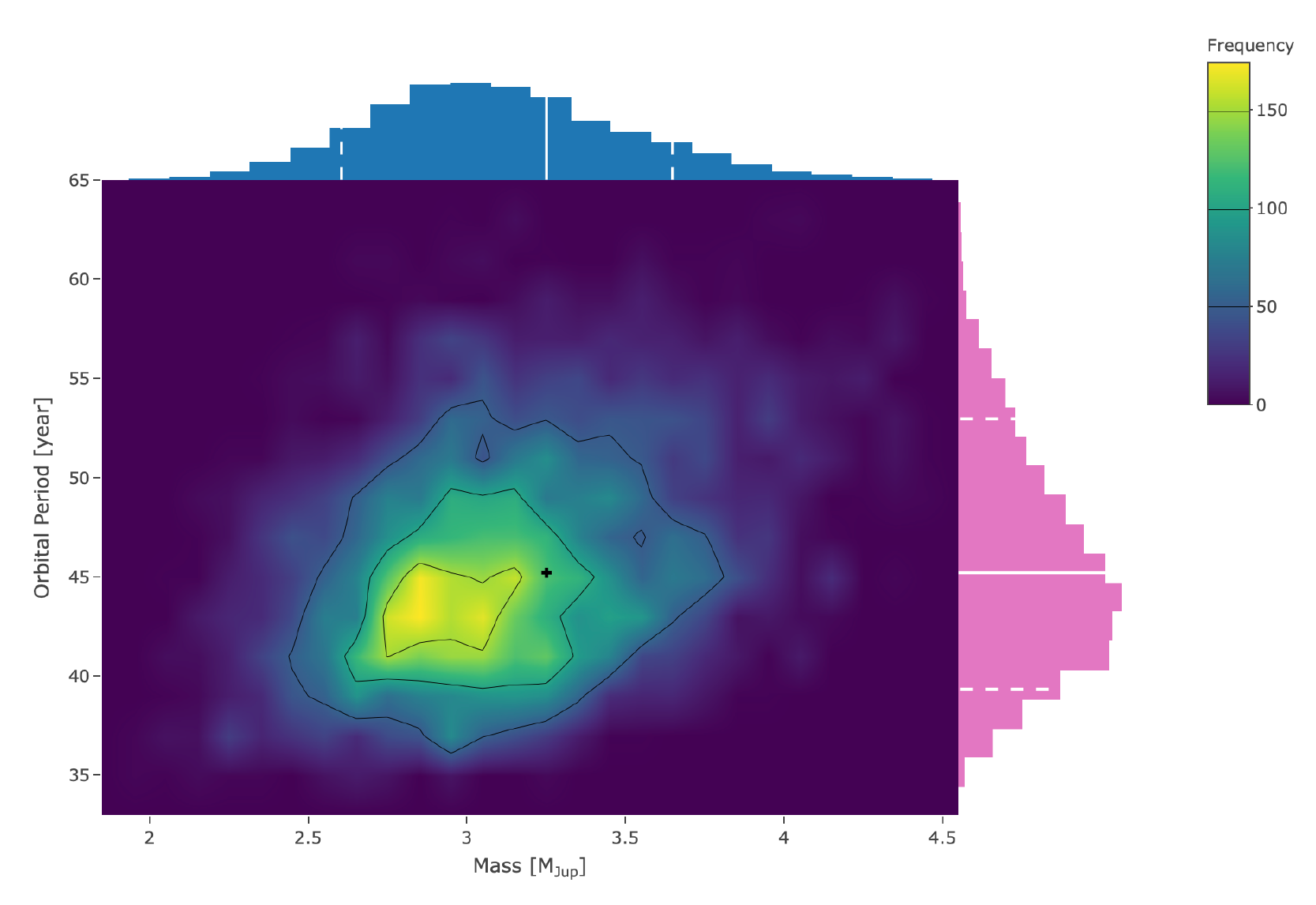}
\caption{Posterior distribution of the mass and orbital period of $\epsilon$ Indi A b. The contour is calculated from a subsample drawn by the MCMC chains from the posterior distribution. Three contour levels corresponding to one, two and three sigma levels are shown by black lines. The MAP value (white solid lines) and the 10\% and 90\% quantiles (white dashed lines) of the orbital period and planetary mass are shown in the histograms of the mass and orbital period. The black cross in the contour denotes the MAP solution. This MAP solution differs from the mean solution because the distributions of the mass and orbital parameters are not Gaussian and are asymmetric.}
\label{fig:mp}
\end{figure*}

\section{UVES RV data }
\begin{table}
  \caption{UVES RV data. }
  \label{tab:data}
  \centering
  \begin{tabular}{c*2{c}}
\hline\hline
    BJD&RV&RV error\\
    (day)&(m/s)&(m/s)\\
        \hline
2453272.49027&-0.09&0.85 \\
2453272.4912&-1.57&0.89 \\
2453272.49202&0.65&0.81 \\
2453272.49285&1.32&0.74 \\
2453272.49369&-1.6&0.87 \\
2453272.49452&-0.8&0.79 \\
2453272.49536&0.4&0.88 \\
2453272.49619&0.85&0.96 \\
2453272.49701&-0.23&0.84 \\
2453272.49783&-1.44&0.89 \\
2453272.49866&-1.4&0.91 \\
2453272.49983&-5.57&1.04 \\
2453272.50075&-0.36&0.75 \\
2453272.50152&-0.57&0.75 \\
2453272.50253&-1.15&0.7 \\
2453272.5033&-0.34&0.76 \\
2453272.50408&-1.3&0.83 \\
2453272.50578&-2.58&0.63 \\
2453272.50657&-1.21&0.58 \\
2453272.5074&-1.55&0.58 \\
2453272.50831&-0.68&0.59 \\
2453272.50916&-0.12&0.57 \\
2453272.50999&0.4&0.61 \\
2453272.51084&-0.79&0.65 \\
2453272.51165&-1.43&0.65 \\
2453272.51249&-0.94&0.62 \\
2453272.51329&0.48&0.67 \\
2453272.51519&-0.96&0.64 \\
2453272.51597&-0.66&0.61 \\
2453272.51676&-1.59&0.6 \\
2453272.51754&-1.25&0.6 \\
2453272.5183&-2.49&0.61 \\
2453272.51908&-0.95&0.58 \\
2453272.51983&-0.43&0.64 \\
2453272.52057&-1.63&0.62 \\
2453272.5213&-0.59&0.64 \\
2453272.52203&-3.72&0.66 \\
2453272.52391&-1.05&0.7 \\
2453272.52471&-1.57&0.66 \\
2453272.5255&-0.93&0.67 \\
2453272.5263&-0.07&0.67 \\
2453272.52711&-1.36&0.65 \\
2453272.52792&-1.95&0.66 \\
2453272.52869&-1.58&0.64 \\
2453272.52947&-1.18&0.71 \\
2453272.53024&-1.39&0.6 \\
2453272.53101&-0.96&0.67 \\
2453272.53286&0&0.77 \\
2453272.5336&-2.54&0.8 \\
2453272.53432&-1.16&0.77 \\
2453272.53505&-0.83&0.79 \\
2453272.53917&0.09&1.01 \\
2453272.53987&-0.25&0.91 \\
    \hline
      \end{tabular}
  \end{table}

  \begin{table}
    \caption{Table \ref{tab:data} continue. }
    \label{tab:data2}
      \centering
  \begin{tabular}{c*2{c}}
\hline\hline
    BJD&RV&RV error\\
    (day)&(m/s)&(m/s)\\
    \hline
    2453272.54076&-0.48&0.83 \\
2453272.54151&-0.47&0.65 \\
2453272.54228&-0.94&0.69 \\
2453272.54304&-0.24&0.69 \\
2453272.54382&0.26&0.63 \\
2453272.54464&-1.43&0.64 \\
2453272.54546&-0.25&0.66 \\
2453272.54626&0.3&0.62 \\
2453272.54706&-2.44&0.67 \\
2453272.54782&-0.13&0.72 \\
2453272.55005&-0.15&0.8 \\
2453272.55081&-1.1&0.61 \\
2453272.55158&-1.15&0.63 \\
2453272.55236&0.1&0.66 \\
2453272.55314&-0.19&0.63 \\
2453272.55393&-1.02&0.62 \\
2453272.55473&-1.84&0.56 \\
2453272.55552&-2.06&0.71 \\
2453272.55632&-0.74&0.62 \\
2453272.55718&-2.83&0.61 \\
2453272.55823&-1.37&0.56 \\
2453272.55997&-1.39&0.62 \\
2453272.56082&-2.43&0.63 \\
2453272.56168&-0.22&0.64 \\
2453272.56255&-1.36&0.59 \\
2453272.56337&-0.14&0.69 \\
2453272.56418&-3.49&0.7 \\
2453272.56499&-0.85&0.68 \\
2453272.5658&-0.62&0.68 \\
2453272.56662&-1.93&0.61 \\
2453272.56745&-0.03&0.64 \\
2453272.56828&-2.14&0.62 \\
2453272.5702&-2.37&1.13 \\
2453272.57094&0.26&0.77 \\
2453272.57173&-1.13&0.65 \\
2453272.57254&-1.59&0.67 \\
2453272.57334&-1.85&0.67 \\
2453272.57413&0.06&0.73 \\
2453272.57492&-2.69&0.69 \\
2453272.57572&-0.67&0.67 \\
2453272.57652&-1.71&0.69 \\
    2453272.57733&-1.24&0.7 \\
        2457905.83292&7.65&0.93 \\
2457905.839&6.92&0.94 \\
2457905.8412&7.45&0.93 \\
2457905.84344&5.56&0.86 \\
2457905.84566&7.99&0.88 \\
2457905.86181&6.97&0.93 \\
2457905.86267&8.5&0.95 \\
2457905.86353&9.13&0.93 \\
2457905.86439&8.53&0.96 \\
2457905.86689&7.93&0.98 \\
2457905.86774&8.08&0.96 \\
2457905.86861&7.65&0.96 \\
2457905.86947&6.17&1.01 \\
2457905.87185&8.67&0.92 \\
2457905.8727&9.02&0.91 \\
   \hline
      \end{tabular}
    \end{table}

      \begin{table}
    \caption{Table \ref{tab:data} continue. }
    \label{tab:data3}
      \centering
  \begin{tabular}{c*2{c}}
\hline\hline
    BJD&RV&RV error\\
    (day)&(m/s)&(m/s)\\
    \hline

2457905.87356&7.78&0.95 \\
2457905.87442&8.69&0.92 \\
2457905.8768&7.64&0.98 \\
2457905.87766&6.51&0.95 \\
2457905.87852&6.94&0.89 \\
2457905.87939&8.87&0.9 \\
2457905.88175&9.46&0.92 \\
2457905.88262&8.36&0.94 \\
2457905.88347&7.95&0.92 \\
2457905.88433&7.99&0.9 \\
2457905.8868&6.58&1.02 \\
2457905.88765&6.43&0.94 \\
2457905.88852&7.75&1.04 \\
2457905.88938&5.49&0.97 \\
2457905.89185&7.1&0.95 \\
2457905.89271&10.2&0.94 \\
2457905.89357&8.43&0.95 \\
2457905.89443&7&0.96 \\
2457905.8969&8.72&0.96 \\
2457905.89775&8.33&0.91 \\
2457905.89861&7.88&0.97 \\
2457905.89948&7.74&0.93 \\
2457993.67726&5.97&0.84 \\
2457993.67812&6.95&0.91 \\
2457993.67898&7.09&0.86 \\
2457993.67984&7.72&0.87 \\
2457993.68162&7.57&0.86 \\
2457993.68247&7.71&0.93 \\
2457993.68334&8.27&0.94 \\
2457993.6842&8.63&0.89 \\
2457993.68594&7.33&0.88 \\
2457993.68679&6.02&0.96 \\
2457993.68765&8.74&0.92 \\
2457993.68851&6.31&0.89 \\
2457993.69038&6.71&0.92 \\
2457993.69124&7.19&0.91 \\
2457993.6921&8.14&0.87 \\
2457993.69296&6.28&0.95 \\
2457993.69527&6.47&0.93 \\
2457993.69613&4.92&0.9 \\
2457993.69699&6.77&0.98 \\
2457993.69786&7.28&0.93 \\
2457993.69971&7.86&0.9 \\
2457993.70057&7.29&0.87 \\
2457993.70144&7.69&0.82 \\
2457993.70229&7.28&0.88 \\
2457993.70522&6.38&0.89 \\
2457993.70608&7.32&0.93 \\
2457993.70693&5.77&0.88 \\
2457993.7078&6.36&0.96 \\
2457993.70968&6.6&0.87 \\
2457993.71054&6.81&0.92 \\
2457993.7114&7.27&0.88 \\
2457993.71226&7.82&0.98 \\
   \hline
      \end{tabular}
  \end{table}
 \bibliographystyle{mn2e}
\bibliography{nm}
% \end{CJK*}
\end{document}